\documentclass[a4paper,11pt]{article}
\pdfoutput=1 

\usepackage{jheppub} 

\usepackage[T1]{fontenc} 

\newcommand*\longhookrightarrow{\ensuremath{\lhook\joinrel\relbar\joinrel\rightarrow}}

\title{\boldmath Towards holographic higher-spin interactions: Four-point functions
and higher-spin exchange}

\author[a]{X. Bekaert,}
\author[b]{J. Erdmenger,}
\author[c]{D. Ponomarev,}
\author[b]{and C. Sleight}

\affiliation[a]{Laboratoire de Math\'{e}matiques et Physique Th\'{e}orique,\\Unit\'{e} Mixte de Recherche 7350 du CNRS,\\ F\'{e}d\'{e}ration de Recherche 2964 Denis Poisson Country,\\Universit\'{e} Fran\c{c}ois Rabelais, Parc de Grandmont, 37200 Tours, France}
\affiliation[b]{Max-Planck-Institut f\"{u}r Physik (Werner-Heisenberg-Institut),\\F\"{o}hringer Ring 6, D-80805 Munich, Germany}
\affiliation[c]{Arnold Sommerfeld Center for Theoretical Physics,\\Ludwig-Maximilians University Munich,\\Theresienstr. 37, D-80333 Munich, Germany}

\emailAdd{xavier.bekaert@lmpt.univ-tours.fr}
\emailAdd{jke@mpp.mpg.de}
\emailAdd{dmitry.ponomarev@physik.uni-muenchen.de}
\emailAdd{csleight@mpp.mpg.de}

\abstract{Within holography, we calculate the contribution of an arbitrary spin-$s$ gauge boson exchange in AdS$_{d+1}$ to the four-point function with scalar operators on the boundary. As an important ingredient, we first compute the complete bulk-to-bulk propagators for massless bosonic
 higher-spin fields in the metric-like formulation, in any dimension and in various gauges.
The split representation of the bulk-to-bulk propagators in terms of bulk-to-boundary propagators allows to present the higher-spin exchange diagram in the form of
a conformal partial wave expansion. Our results provide a step towards the larger goal of the holographic reconstruction of bulk interactions, and of clarifying bulk locality.}

\begin{document} 
\maketitle
\flushbottom

\section{Introduction}

Bulk locality remains one of the most important and elusive properties of the anti-de Sitter / conformal field theory (AdS/CFT) correspondence \cite{Maldacena:1997re,Gubser:1998bc,Witten:1998qj,Aharony:1999ti}. 
This property is expected to hold in the usual regime where the duality is tested: when the AdS radius is large compared to the Planck and string lengths which, on the CFT side, corresponds to a large-$N$ expansion
and a gap in the spectrum of anomalous dimensions.
The latter two properties were argued to provide necessary and sufficient conditions for a CFT to possess a local bulk dual
\cite{Heemskerk:2009pn,Fitzpatrick:2010zm,ElShowk:2011ag} (a third condition was added in \cite{Fitzpatrick:2012cg}). 
On the other hand, the conjectured duality \cite{Sezgin:2002rt,Klebanov:2002ja} between Vasiliev's higher-spin gravity \cite{Vasiliev:1990en,Vasiliev:1992av,Vasiliev:2003ev} and a vector model at a free or critical fixed point provides a convenient playground for probing deep issues, such as bulk locality, in holography. This is because it holds in a regime where in principle both sides are calculable, in contrast to that of the standard AdS/CFT correspondence described above. A particular example that should be tractable is the explicit computation of the quartic vertex in higher-spin gravity, matching
the four-point correlator of the free CFT. The result may shed some light on the issue of bulk locality in higher-spin holography, and the present paper aims to prepare the technical tools for attacking
the above concrete match.

In the recent years, there has been great progress connecting CFT correlation functions to scattering processes in AdS spacetime
\cite{Penedones:2010ue,Fitzpatrick:2011hu,Fitzpatrick:2011dm,Costa:2012cb}. This progress was based on the technology of Mellin amplitudes 
\cite{Mack:2009mi,Fitzpatrick:2011ia,Paulos:2011ie,Nandan:2011wc}.
The programme of reconstructing bulk theories from their dual CFTs via the rewriting of Mellin amplitudes as Witten diagrams
appears to apply to a large class of strongly-coupled CFTs. Unfortunately it does not directly\footnote{Technically, this can be seen by computing the correlation function of four
single-trace scalar operators in the free $O(N)$ vector model via Wick contraction. This gives 
\begin{equation}
F(u,v)\propto u^\Delta+(u/v)^\Delta+u^\Delta(u/v)^\Delta
\label{Fuv}
\end{equation}
where $\Delta=d-2$ is the scaling dimension of the single-trace scalar operator and $F(u,v)$ is the factor depending on the two cross ratios that is not fixed by conformal symmetry in the four-scalar correlation function. 
Following the prescription \cite{Mack:2009mi}, the corresponding 4-point Mellin amplitude should be proportional to the Mellin transform of $F(u,v)$ over both variables. However, the former is \textit{not} well-defined since
the function \eqref{Fuv} is a sum of products of powers of $u$ and $v$, while power functions do not admit a Mellin transform.\label{footn}} apply to the simplest example of weakly-coupled CFTs: 
free scalar fields. Nevertheless, a putative bulk dual appear to exist in the form of Vasiliev higher-spin gravity.
At a conceptual level, the fact that free CFTs fall outside the scope of the above holographic reconstruction programme can be seen by inspecting the set of necessary and sufficient conditions, proposed in
\cite{Fitzpatrick:2012cg}, for a CFT to possess an ``AdS effective field theory'' dual: these conditions are not satisfied by free large-$N$ CFTs. Firstly, the spectra of free CFTs contain a gapless infinite set of
single-trace primary operators. Secondly and more importantly, their Mellin amplitudes may not be bounded by a polynomial of Mellin space variables, actually they may even not be defined at all
(\emph{cf.} footnote \ref{footn}). In bulk terms, this translates into the fact that, first, the spectrum of bulk fields contains an infinite tower of massless fields with unbounded spin and, 
second, that the bulk theory does not admit an expansion in (non-negative) powers of the cosmological constant. These properties are perfectly consistent with key features of Vasiliev 
theory (see \emph{e.g.} \cite{Bekaert:2010hw} for a nontechnical review), and explain which assumptions of the Weinberg low-energy theorem\footnote{Which prevents long-range higher-spin exchanges.} generalised to AdS \cite{Fitzpatrick:2012cg} are circumvented by higher-spin gravity.
Nevertheless, a tantalising open question remains: what is precisely the status of locality in higher-spin theories?
 
Clarifying this issue is of fundamental importance, since locality is one of the core properties of fundamental field theories. But even in the familiar setting of QFT, the issue is a subtle one since for instance the Wilsonian view on QFT is based on effective field theories, which are quasilocal. This is in the sense that
they possess a perturbative expansion 
(in powers of fields and their derivatives) where each individual term in the total Lagrangian is local, though the total number of derivatives may be unbounded in the
full series.\footnote{In more mathematical terms, a ``quasi'' (or ``perturbatively'') local functional is the spacetime integral of a
power series on the (infinite jet) space spanned by the fields and all their possible derivatives.}
Indeed, a natural candidate of a suitably enlarged definition of ``AdS effective field theory'' (in order to possibly contain Vasiliev theory as a paradigmatic example of bulk dual to a free CFT) is quasilocality.
An immediate proviso is the fact, often emphasised by Vasiliev, that there is no well-defined derivative expansion around (A)dS background. 
More precisely, the 
expansion in the number of (covariant) derivatives mixes with the expansion in powers of the cosmological constant, since the commutator of two background covariant derivatives is of the same order as the cosmological constant. 
A second proviso is the fact that higher-spin interactions are weighted by powers of the AdS length. This property is responsible for the absence of a weakly-coupled flat limit.

The ``unfolded'' equations of Vasiliev (see \emph{e.g.} \cite{Bekaert:2005vh,Didenko:2014dwa} for self-contained pedagogical reviews) provide a compact system of equations for which diffeomorphism invariance and 
formal consistency are manifest\footnote{These properties are automatic because Vasiliev equations take the form of a what is 
often called a ``free differential algebra'' in the physics literature.} and whose linearisation is equivalent to a system of Fronsdal's equations \cite{Fronsdal:1978rb,Fronsdal:1978vb} describing an 
infinite tower of massless higher-spin fields.
The price to pay for their concision is that the fields 
appearing in these equations are generating functions for the infinite collection of dynamical fields, together with a plethora of auxiliary fields. 
Indeed, auxiliary fields are introduced at each of the three key steps of Vasiliev's construction: the frame-like formulation (one adds generalised spin-like connections for each metric-like Fronsdal field), 
the unfolding procedure (introduction of an auxiliary field for each on-shell non-trivial derivative of the dynamical fields) and the doubling step
(an expansion in terms of the auxiliary variables $Z$ which allows to reconstruct the interactions). The unfolding procedure seems deeply rooted in the higher-derivative nature of higher-spin interactions and symmetries.
This procedure is also very natural from a mathematical\footnote{The philosophy underlying Vasiliev's unfolding procedure is similar to Cartan's prolongation method, where one
replace differential equations by a system of algebraic equations on the jet space of fields and their derivatives.} viewpoint and should guarantee the absence of strong
non-localities (such as inverse powers of the wave operator).

Nevertheless, basic issues such as the precise form that the expected quasilocality takes in higher-spin-gravity remain technically difficult to address in the unfolded
formulation, due to the dressing of each dynamical field with an infinite collection of auxiliary fields. 
For that reason, we prefer to address the issue in the more pedestrian context of the Noether approach for metric-like fields.
A concrete well-posed question one might like to address is whether the contact terms relevant in a Witten diagram with prescribed external legs are local or not. 
This is known to be true at the level of three-point functions\footnote{This is a straightforward consequence on known bounds on the total number of derivatives of nontrivial cubic vertices on
AdS for any given triplet of spins, see \emph{e.g.} \cite{Joung:2012fv,Joung:2012hz,Boulanger:2012dx} for recent results. Bulk locality of three-point functions was recently discussed in \cite{Sarkar:2014dma}.\label{footn2}} but
we are unaware of any result in this direction for higher-point functions.\footnote{However, see \cite{Vasiliev:1989yr,Metsaev:1991mt,Fotopoulos:2010ay,Polyakov:2010sk,Taronna:2011kt,Dempster:2012vw,Florakis:2014kfa} on quartic interactions.} As emphasised in \cite{Bekaert:2010hw}, this issue is important because any cubic vertex which is gauge invariant till this order can be consistently completed by non-local higher-order (quartic, etc) vertices \cite{Barnich:1993vg}. It is the assumption of (quasi)locality that imposes very strong constraints on the set of possibilities.

To tackle this question, our strategy is to look to the AdS/CFT correspondence for some assistance. We work in the context of bosonic higher-spin gravity on AdS$_{d+1}$ ($d>2$),\footnote{For AdS$_3$/CFT$_2$ higher-spin holography,
see \cite{Gaberdiel:2012uj} and references therein.}
which for un/broken higher-spin symmetry is conjectured 
to be dual to a CFT$_d$ of massless scalars in the vector representation of an internal symmetry group at the free/critical fixed point. The duality has already passed remarkable tests at tree level 
for 3-point Witten diagrams \cite{Giombi:2009wh,Giombi:2010vg} (see also \cite{Giombi:2012ms} for a review), and at one-loop level for vacuum free energies and Casimir energies \cite{Giombi:2013fka,Giombi:2014iua,Giombi:2014yra,Beccaria:2014xda}. Moreover, the $n$-pt correlation 
functions of the free CFT$_3$ have been obtained directly in the unfolded formulation via suitable traces in the auxiliary twistor space \cite{Colombo:2012jx,Didenko:2012tv,Gelfond:2013xt}. Within the
metric-like formulation, tests of the correspondence at the quartic level (\emph{i.e.} for four-point functions) are yet to be achieved, as they would require the knowledge of quartic higher-spin interactions contributing to 
the total amplitude. As mentioned above, these are currently unknown in the metric-like formulation. On the other hand, in possession of the CFT result (which is relatively easy to compute) and the ability to calculate the remaining contributing Witten 
diagrams that involve lower-order (known) interactions (\emph{cf.} refs in footnote \ref{footn2}.), one should instead be able to apply the correspondence to infer the form of the quartic vertex and hence establish whether or not they are local.

\begin{figure}[t]
 \centering
\includegraphics[width=0.9\linewidth]{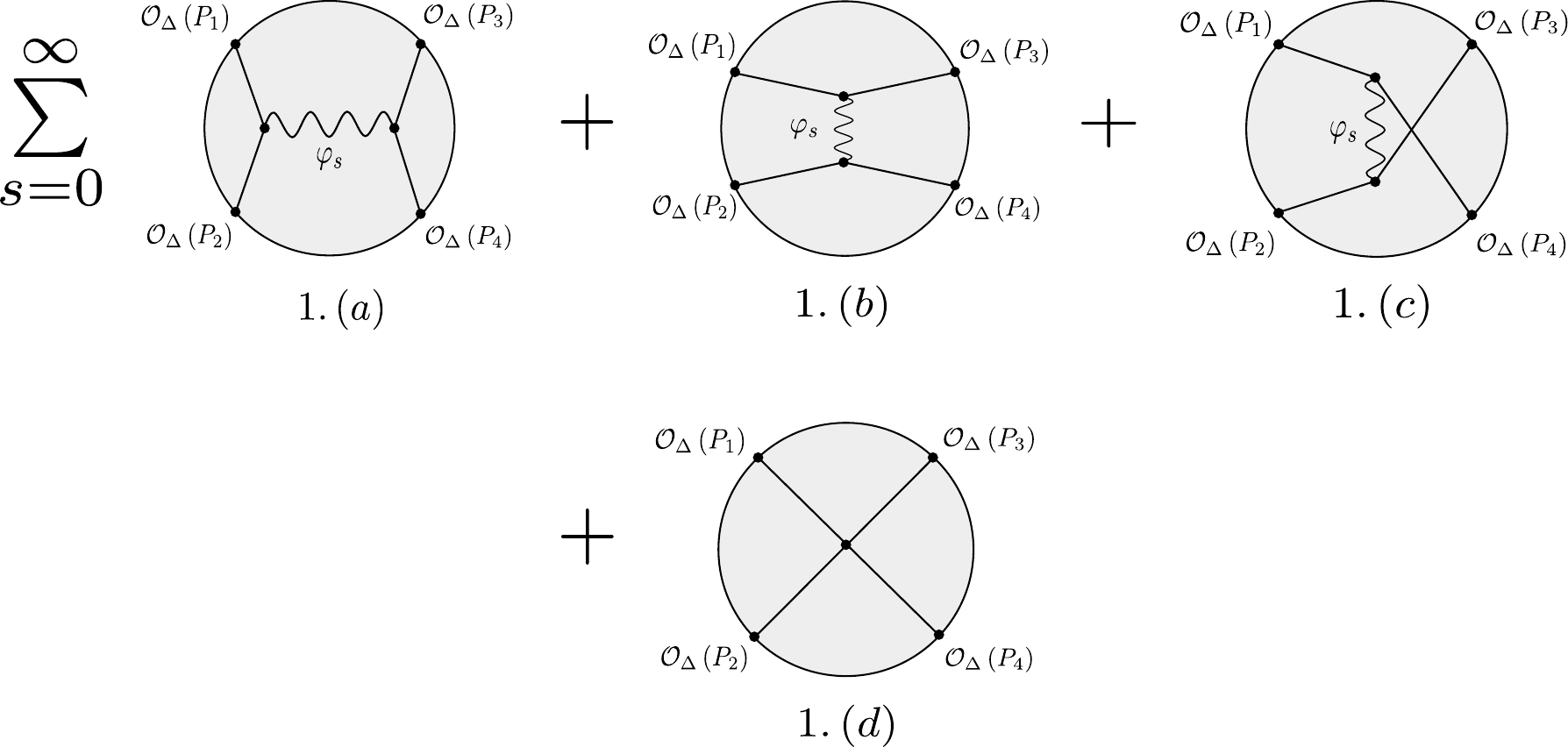}
 \caption{Total four-point Witten diagrams contributing to the bulk computation of the four-point function of the scalar singlet bilinear operator $\mathcal{O}_{\Delta}$ in the $O\left(N\right)$ vector model. Here, $\Delta$ is the dimension of the bulk scalar dual to $\mathcal{O}_{\Delta}$, and the $P_i$ are fixed points on the boundary of AdS$_{d+1}$. In the present paper, we compute diagram 1.(a) (the ``s-channel'') for a spin-$s$ gauge boson $\varphi_s$. Computations for the remaining exchange channels 1.(b) and 1.(c) follow in the same way. Diagram 1.(d) illustrates the quartic scalar contact interaction.}
\label{fig:total4pt}
\end{figure}

The most natural place to begin this endeavour is be to simply consider the four-point function of the scalar singlet bilinear operator in the $d$-dimensional free $O\left(N\right)$ vector model, which is proposed to 
be dual to the minimal higher-spin theory on AdS$_{d+1}$ containing gauge fields of all even spins $s = 2, 4, ...$ and a single real scalar field ($s=0$). We work in this minimal framework in the view that our
results can easily be extended to the non-minimal theory of all integer spins  and a complex boundary scalar, and possible generalisations thereof. Within conformal field theory, the scalar four-point function is very 
straightforward to work out via Wick contractions (and is essentially given in footnote \ref{footn}). The Witten diagrams entering the holographic computation are exchanges of gauge fields of all even integer spin between two pairs of 
the real bulk scalar, and of course the quartic scalar contact interaction we seek. These are displayed in figure \ref{fig:total4pt}. In the present paper, we compute the exchange\footnote{In the AdS/CFT literature, four-point exchange computations of lower spin ($s \leq 2$) fields were computed in \cite{Freedman:1998bj,Chalmers:1998wu,Liu:1998ty,DHoker:1998gd,D'Hoker:1998mz,Liu:1998th,Chalmers:1999gc,D'Hoker:1999pj}. In the context of higher-spins, see \cite{Francia:2007qt,Francia:2008hd,Sagnotti:2010jt,Costa:2014kfa}.} of a given massless spin-$s$ gauge field between external scalars of arbitrary mass in AdS$_{d+1}$, laying down some 
of the ground work for the eventual extraction of the quartic vertex.

In order to achieve this goal, certain technical hurdles first need to be overcome. For example, the explicit forms of the bulk-to-bulk propagators of the massless bosonic spin-$s$ fields in the metric-like formalism need
to be established. Scalar and spinor propagators date back to the old literature on AdS field theory \cite{Fronsdal:1965zzb,Burgess:1984ti,Inami:1985wu,Burges:1985qq}, while for the massive and massless spin-1 fields 
they were obtained in \cite{Allen:1985wd}. More recently, massive and massless spin-2 propagators were established in \cite{Liu:1998ty,DHoker:1999jc} and \cite{Naqvi:1999va}, where in the latter propagators for $p$-forms were 
also derived. Spin-2 propagators were also established for de Sitter space in \cite{Allen:1986tt,Turyn:1988af,Antoniadis:1986sb,Gabriel:1996iy}. Unfortunately, the methods employed in these works become intractable when applied in the hope of deriving analogous results for the
higher-spin massless propagators. However, recent applications of harmonic analysis in AdS space have been successful in determining the traceless part of bulk-to-bulk propagators for massive bosonic fields 
of arbitrary integer spin \cite{Costa:2014kfa}.\footnote{For previous literature on higher-spin bulk-to-bulk propagators, see: \cite{Manvelyan:2005fp,Francia:2008hd,Manvelyan:2008ks,Mkrtchyan:2010pp,Didenko:2012vh}.} In the present paper, we adapt these methods to establish the complete off-shell form of the metric-like massless bosonic higher-spin bulk-to-bulk propagators,
in arbitrary dimensions and in various gauges. The possession of the propagators in multiple gauges allows us to check the consistency of our results at various steps along the way.

\begin{figure}[t]
 \centering
\includegraphics[width=0.9\linewidth]{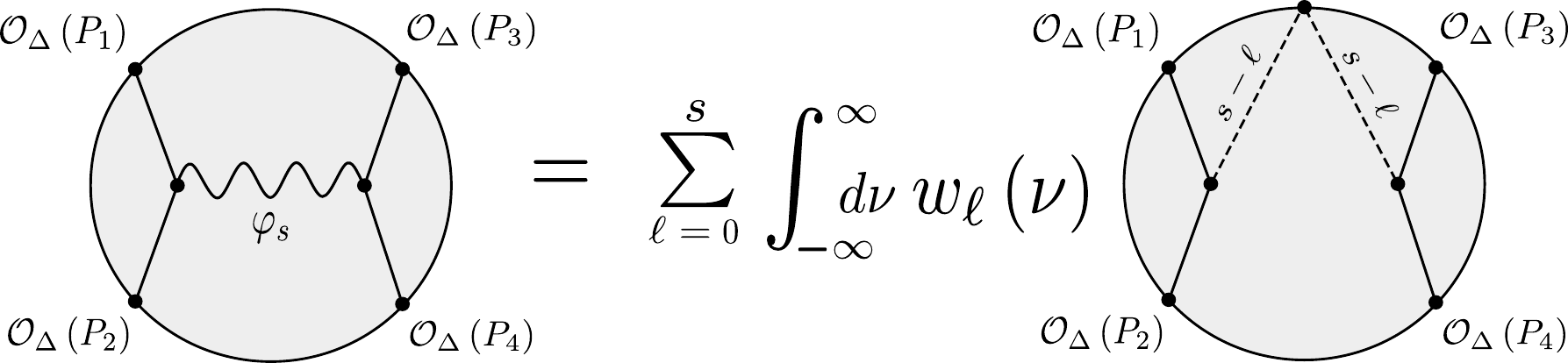}
 \caption{Use of the split representation for the bulk-to-bulk propagators in the exchange results in a decomposition in terms of products of three-point Witten diagrams, involving a pair of real scalar fields and a field whose spin, $s-\ell$, is summed over. The common boundary point of the three-point functions is integrated over. This is to be compared with the definition \eqref{cpwe} of the conformal partial wave expansion in conformal field theory.}
\label{fig:factor}
\end{figure}

A particular virtue of the harmonic analysis is that the resultant form of the bulk-to-bulk propagators 
admits a \emph{split representation}, in which they can be expressed as a sum of integral 
products of two bosonic bulk-to-boundary propagators \cite{Leonhardt:2003qu,Leonhardt:2003sn}.\footnote{The split representations of scalar, spin-1 and spin-2 propagators can be found in \cite{Penedones:2010ue,Paulos:2011ie,Balitsky:2011tw,Costa:2014kfa}. Note that the discussion generated in the literature by the split form of the graviton propagator \cite{Balitsky:2011tw,Giecold:2012qi} was recently addressed in \cite{Costa:2014kfa}.} As a consequence, their use in the exchange computation causes the amplitude to decompose into products of three-point functions (see figure \ref{fig:factor}), for which effective methods to compute are already known.
This decomposition of the exchange amplitude is reminiscent of the conformal partial wave expansion in conformal field theory. Drawing on this similarity, we express our results for the exchange computation such that, when supplemented with the results from
the other exchange channels and summing the contributions from each spin (figure \ref{fig:total4pt}), they can be directly compared with the analogous form of CFT result.\footnote{See for example \cite{Leonhardt:2002sn,Diaz:2006nm}, and the ensuing discussions.} Our computations are greatly facilitated by working in the ambient formalism, which is particularly convenient when addressing higher-spin (AdS or conformal) fields. It has been applied in many contexts, for example in \cite{Biswas:2002nk,Mikhailov:2002bp,Bekaert:2010hk,Grigoriev:2011gp,Joung:2012fv,Didenko:2012vh,Joung:2012hz,Costa:2014kfa}.

In this holographic reconstruction of bulk quartic vertices, there are also certain subtleties regarding cubic interactions that need to be confronted, which ultimately influence the cubic vertices entering the exchanges. At the cubic level, there is a degree of arbitrariness
provided by ``trivial'' vertices that vanish on the free-shell. In the language of conserved currents (since any cubic interaction
involving two scalar fields can be expressed in terms of a coupling between a gauge
field and a current), these correspond to ``improvements'' of genuine Noether currents.\footnote{The relevance of such terms was already emphasised in \cite{Boulanger:2008tg} in the particular case of the ``Born-Infeld tail''.\label{BItail}} In the context of the computation of the exchange diagrams, this ambiguity translates into the question of how such terms should enter in the cubic vertices. In the present paper, we also briefly study the effect of improvement terms in the four-point exchange. These are usually neglected in studies of cubic vertices, however we find that such trivial vertices in fact enter non-trivially in exchange computations in the present context. Further, their presence in the exchange diagrams generates quartic contact terms, such that the original exchange amplitude decomposes into an exchange governed by the on-shell part of the vertices, plus a quartic scalar contact diagram. This highlights a collaboration between quartic and trivial cubic vertices, and we briefly comment on the implication for the holographic study of quartic contact interactions in the conclusion.

\subsubsection*{Summary of results}

The road towards the results for the exchange computation in this paper is a long one, establishing the necessary supplementary new results and making numerous pit stops to introduce the required existing results along the way. 
In part, the reason for this was to make the presentation self contained, and we therefore for convenience briefly summarise our main results below.

In section \ref{Sec:Propagators}, we use harmonic analysis in anti-de Sitter space to derive, in arbitrary dimensions and in various gauges, the complete off-shell form of the bulk-to-bulk propagators
for massless bosonic spin-$s$ fields governed by Fronsdal's equations. This is carried out in the following gauges: de Donder gauge (eq. \eqref{tracelesspartdeDonder}, section \ref{Sec:Propagators;Subsec:deDonder}),
a traceless gauge (eq. \eqref{tlsplit}, section \ref{Sec:Propagators;Subsec:tracelessgauge}), and a gauge which we refer to as the ``manifest trace gauge'' as it expresses the propagator in a decomposition that makes 
manifest the metric dependence (eq's. \eqref{eq:ManifestTrace} and \eqref{answertrgauge}, section \ref{Sec:Propagators;Subsec:ManifestTrace}).

In section \ref{Sec:FourPointExchange}, we then compute the four-point exchange of a single spin-$s$ gauge boson between pairs of the real scalar in the minimal
bosonic higher-spin theory on AdS$_{d+1}$ for the ``s-channel'' -- figure \ref{fig:total4pt}. (a). The computations for the other two channels follow in the same way. This is done in 
the traceless gauge (eq. \eqref{pw}, section \ref{Sec:FourPointExchange;Subsec:TracelessGauge}) and the manifest trace gauge (eq. \eqref{mtpw}, section \ref{Sec:FourPointExchange;Subsec:ManifestTraceGauge}), and the 
results are expressed in the form of a conformal partial wave expansion on the boundary. We check that the computations in both gauges are in agreement.

\subsubsection*{Outline of paper}

We begin by introducing the ambient formalism in section \ref{sec:ambient}, where both bulk AdS isometries and boundary conformal symmetries are manifest. In particular, we review the material necessary for handling symmetric tensor fields on AdS and its boundary in a compact way.

In section \ref{Sec:Propagators}, the metric-like bulk-to-bulk propagators for symmetric tensor gauge bosons on AdS$_{d+1}$ of any spin are determined in three distinct gauges. To do so, in section \ref{Subsec:Fronsdal} we review the metric-like formulation of massless bosonic higher-spin fields in anti-de Sitter space, and specify the equation that the corresponding bulk-to-bulk propagators should satisfy. Solving the propagator equations is facilitated by working in a basis of AdS harmonic functions, which review in section \ref{Sec:Propagators;Subsec:Split}. With this, we then proceed to solve for the propagators in de Donder gauge in section \ref{Sec:Propagators;Subsec:deDonder}, a traceless gauge in section \ref{Sec:Propagators;Subsec:tracelessgauge} and what we define as a ``manifest trace gauge'' in section \ref{Sec:Propagators;Subsec:ManifestTrace}.

Before proceeding to compute the exchange, we first review the existing ingredients we use along the way. We begin with the ambient representation of AdS bulk-to-boundary propagators for symmetric tensor fields in section \ref{sec:3pt}. Then, in section \ref{subsec:split} we introduce the split representation of the derived bulk-to-bulk propagators, in which they are expressed as sums of integral products of pairs of the aforementioned bulk-to-boundary propagators. Further, we discuss the ensuing decomposition of exchange amplitudes, and recall ambient methods to evaluate the resulting three-point bulk integrals. Parallels with the conformal partial wave expansion in CFT are then drawn. We establish the explicit form of the cubic vertex we use between the two bulk scalars and the exchanged higher-spin gauge field in section \ref{Sec:FourPointExchange;Subsec:Currents}.

Finally, in the traceless gauge in section \ref{Sec:FourPointExchange;Subsec:TracelessGauge} and in the manifest trace gauge section \ref{Sec:FourPointExchange;Subsec:ManifestTraceGauge}, we compute the four-point exchange of a single spin-$s$ gauge boson between two pairs of real scalars in AdS$_{d+1}$. In section \ref{subsec:checks} we verify for explicit examples that the exchange computations in the two different gauges are consistent with each other.
We then discuss the subtleties related to improvements of bulk currents in the last section \ref{subsec:improvements}.

We briefly recapitulate our results in section \ref{sec:concl}, and list the issues that remain to be addressed in order to concretely extract the quartic scalar vertex from the holographic correspondence via the scalar singlet bilinear four-point function of the free vector model.

Some technical points have been relegated to appendices: Appendix \ref{appendix:operators} contains various useful formulae on ambient tensors, appendix
\ref{app:tracelessexch} details the computation of the exchange amplitude in the traceless gauge, while appendices \ref{apptraceofcur} and
\ref{apptraceofcur1} provide the formulae relating respectively the single and multiple traces of bulk conserved currents to lower rank currents.
\newpage

\section{Ambient space formalism}\label{sec:ambient}
\begin{figure}[h]
 \centering
\includegraphics[width=0.45\linewidth]{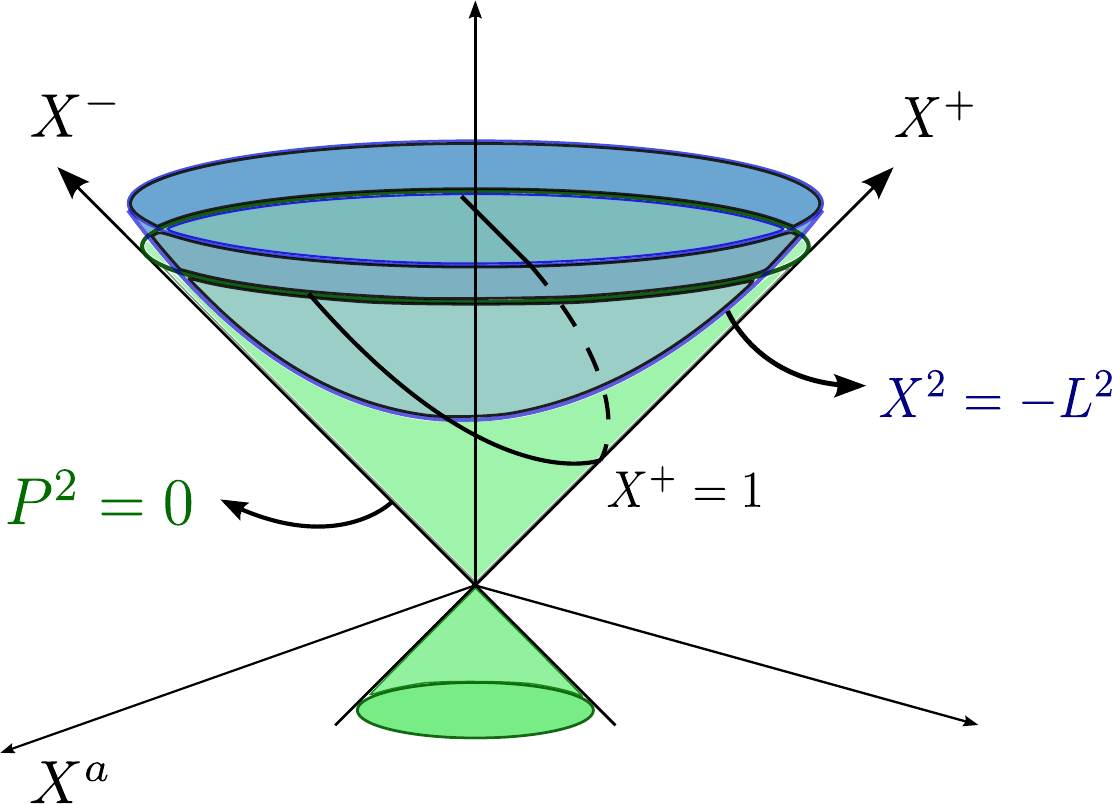}
 \caption{Visualisation of the ambient space with space-like axes $X^a$ $\left(a=1,...d\right)$, and light-like axes $X^+$ and $X^-$. The hypercone $X^2=0$ is in green and the hyperboloid $X^2=-L^2$ is in blue. 
 The paraboloid is the Poincar\'e section obtained by intersecting the light cone with a light-like hyperplane at constant $X^+$, and is illustrated in the figure for $X^+=1$.}
\label{fig:ambient}
\end{figure}

In this paper, we work with fields defined on $\left(d+1\right)$-dimensional Euclidean anti-de Sitter space which will be denoted in the sequel, 
with a slight but standard abuse of terminology, as AdS$_{d+1}$.\footnote{All results can be Wick-rotated to Minkowski signature, with careful treatment of the $i\epsilon$ prescription.} 
In fact, Euclidean anti-de Sitter space is nothing but another name for hyperbolic space. It is often realised as the $\left(d+1\right)$-dimensional Poincar\'e ball, therefore its $d$-dimensional 
conformal boundary $\partial$AdS$_{d+1}$ is topologically a sphere $S^d$.

As was first noted by Dirac \cite{Dirac:1935zz,Dirac:1936fq}, in such a set up it is often useful to make the mutual $SO\left(d+1,1\right)$ symmetry of AdS$_{d+1}$ and its 
conformal boundary manifest, so that the consequential symmetry constraints are manifestly realised. This is particularly effective when considering fields of non-zero spin.
In this formalism, AdS$_{d+1}$ is viewed as a one-sheeted hyperboloid $H_{d+1}$ of curvature radius $L$,
\begin{equation}
H_{d+1}:\quad X^2 = -L^2, \qquad X^{0}> 0, \label{hyperb}
\end{equation}
living in a $\left(d+2\right)$-dimensional flat ambient space $\mathbb{R}^{d+1,1}$. Here we denote the Cartesian coordinates of the ambient space by $X^A$ $\left(A=+,-,1,...,d\right)$, 
and the space itself is endowed with Minkowski metric $\eta_{AB}$ of signature $\left(\;-\;+\;...\;+\;\right)$ which defines
the quadratic form $X^2=\eta_{AB}X^AX^B$.

To be more precise, denoting the \emph{intrinsic} coordinates on $H_{d+1}$ by $x^{\mu}$, we are considering the isometric smooth embedding
\begin{equation}
i\; : \quad H_{d+1} \: \longhookrightarrow \: \mathbb{R}^{d+1,1} : \quad x^{\mu} \longmapsto X^{A}\left(x^{\mu}\right).
\end{equation}
For example, in Poincar\'e coordinates $x^{\mu} = \left(z,\;y^{m}\right)$, $m=0,..,d-1$,
\begin{equation}
X^A=\frac{1}{z}\left(1,\;z^2+y^2,\;y^{m}\right). \label{ppatch}
\end{equation}
Towards the boundary of AdS$_{d+1}$, the hyperboloid asymptotes to the light cone $X^2=0$, and the conformal boundary is identified with the ambient projective cone of light rays. These are 
described by ambient homogeneous coordinates $P^A$, subject to
\begin{equation}
P^2=0, \qquad P \sim \lambda P, \qquad \lambda \neq 0, \label{lightcone}
\end{equation}
where the equivalence relation expresses that one deals with rays.

In the Poincar\'e patch,
\begin{equation}
P^A=\frac{1}{y}\left(1,\;y^2,\;y^{m}\right), 
\end{equation}
which traces out a Poincar\'e section of the light cone along $X^{+}=1$. This set up is summarised in figure \ref{fig:ambient}.

Throughout this paper we use $P^{A}$ for the ambient space representation of points on $\partial$AdS$_{d+1}$, and $X^{A}$ for the ambient AdS$_{d+1}$ coordinates.

The isometry group $SO\left(d+1,1\right)$ acts linearly on the ambient $\mathbb{R}^{d+1,1}$, which gives rise to the isometry group action on AdS$_{d+1}$ and the action of the conformal group on $\partial$AdS$_{d+1}$.
By expressing fields on AdS$_{d+1}$ and its conformal boundary  
in terms of $SO\left(d+1,1\right)$-covariant fields defined in the ambient space, their $SO\left(d+1,1\right)$ symmetry is made manifest. Precisely how this is attained is reviewed in the following sections. 

\subsection{Ambient AdS tensors}
To compute the four-point exchange of arbitrary rank AdS$_{d+1}$ tensors in the ambient framework, it is required to establish how such tensors are represented. A smooth rank-$r$ covariant tensor
field $t_{\mu_1 ...\mu_r}\left(x\right)$ on AdS$_{d+1}$ is represented in ambient space $\mathbb{R}^{d+1,1}$ by a $SO\left(d+1,1\right)$-tensor $T_{A_1...A_r}\left(X\right)$, whose pullback onto the AdS manifold satisfies
\begin{align}
i^{*}\: : \: T_{A_1...A_r}\left(X\right) \: \longmapsto \: t_{\mu_1...\mu_r}\left(x\right) = \frac{\partial X^{A_1}\left(x\right)}{\partial x^{\mu_1}}\, ...\, 
\frac{\partial X^{A_r}\left(x\right)}{\partial x^{\mu_r}} T_{A_1...A_r}\left(X\left(x\right)\right). \label{pullback}
\end{align}
The pullback is surjective, so every such tensor on the AdS manifold has an ambient representative, but it is not injective --- they are not represented uniquely. This is because uplifting to one higher
dimension introduces extra degrees of freedom. Indeed, since for $H_{d+1}$
\begin{equation}
X^2 = -L^2 \: \implies \: \frac{\partial X}{\partial x^{\mu}} \cdot X \: \bigg|_{H_{d+1}}\,= 0, 
\end{equation}
the kernel of the pullback \eqref{pullback} contains ``pure gauge'' tensors with components normal to the AdS manifold, which have no influence in the theory defined 
on AdS.\footnote{For example, for ambient vector fields $V^{A}\left(X\right)$, the kernel is spanned by vector fields of the form $V^{A}\left(X\right)=X^{A}S\left(X\right)$.}

To obtain a unique representation of AdS tensors in ambient space, one needs to eliminate the extra components introduced. This can be achieved by projecting the ambient tangent bundle onto the hyperboloid tangent bundle. In practice, one considers ambient tensors which are tangent to AdS in the sense that they satisfy
\begin{equation}
X^{A_1} T_{A_1...A_r}\bigg|_{H_{d+1}} = 0.
\end{equation}
Explicitly, one can apply the projection operator
\begin{equation}
\mathcal{P}^{B}_{A} = \delta^{B}_{A} + \frac{X_{A} X^{B}}{L^2}, \label{oproj}
\end{equation}
which acts on ambient tensors as
\begin{align}
\left(\mathcal{P} T\right)_{A_1...A_r} \; := \; \mathcal{P}^{B_1}_{A_1}...\mathcal{P}^{B_r}_{A_r}T_{B_1 ...B_r}, \qquad X^{A_i} \left(\mathcal{P} T\right)_{A_1...A_i...A_r}=0.
\end{align}
For example, the intrinsic AdS metric
\begin{align}
g_{\mu \nu} = \frac{\partial X^{A}}{\partial x^{\mu}} \frac{\partial X^{B}}{\partial x^{\nu}} \eta_{AB},
\end{align}
can be represented by the ambient tensor
\begin{align}\nonumber
G_{AB} = \mathcal{P}^{C}_{A}\mathcal{P}^{D}_{B} \eta_{CD} & = \eta_{AB} + \frac{X_{A} X_{B}}{L^2} \\ \nonumber \\  \label{ambmetric}
& =  \eta_{AB} + X_{A} X_{B}.  
\end{align}
In the final equality of the above we set the AdS radius $L=1$, which we adopt throughout the rest of the paper.

It is natural to expect that the ambient representative $\nabla_{A}$ of the covariant derivative $\nabla_{\mu}$ in AdS space is related to taking a partial derivative $\partial_{A}$ in the flat 
ambient space. However care must be taken: Schematically,
\begin{equation}
\label{cpvdercur}
\nabla = {\cal P} \circ \partial \circ {\cal P}
\end{equation}
which ensures that the result is tangent to the AdS manifold, and also crucially that the partial derivative is acting on an object which already represents an AdS tensor. For example,
\begin{equation}
\nabla_{B} T_{A_1...A_r} = \mathcal{P}_{B}{}^{C}\mathcal{P}_{A_1}{}^{C_1}...\mathcal{P}_{A_r}{}^{C_r} \frac{\partial}{\partial X^{C}} \left(\mathcal{P}T\right)_{C_1...C_r}\left(X\right).
\end{equation}
\subsubsection*{Symmetric AdS tensors}
We consider higher-spin fields in AdS$_{d+1}$ represented by symmetric tensors, and often work with tensors which are both symmetric and traceless. In the following, we review how such tensors can be managed in a compact way.

For symmetric tensors, to handle the indices efficiently we can encode them in generating functions. For intrinsic symmetric tensors, 
\begin{equation}
t_{\mu_1...\mu_r}\left(x\right) \: \longrightarrow \: t\left(x,u\right) = \frac{1}{r!} t_{\mu_1...\mu_r}\left(x\right) u^{\mu_1}...u^{\mu_r},
\end{equation}
 where we have introduced the constant auxiliary variable $u^{\mu}$.

 For the corresponding ambient tensors, tangent to AdS$_{d+1}$ 
 \begin{equation}
 T_{A_1...A_r}\left(X\right) \: \longrightarrow \: T\left(X,U\right) = \frac{1}{r!} T_{A_1...A_r}\left(X\right) U^{A_1}...U^{A_r}, \quad \text{with}\quad X \cdot U = 0. 
 \end{equation} 
 Here, the constrained ambient auxiliary vector $U^A$ ensures that we are working modulo components which drop out after projection \eqref{oproj} onto the tangent space of $H_{d+1}$, via $X \cdot U = 0$.

In sum, the ambient representative $T_{A_1...A_r}\left(X\right)$ of a symmetric tensor $t_{\mu_1...\mu_r}\left(x\right)$ on AdS$_{d+1}$ can be fully encoded in a polynomial $T\left(X,U\right)$ defined on 
  the submanifold $X^2+1=0=X \cdot U$.
\subsubsection*{Symmetric and traceless AdS tensors}
\label{sec:sandtads}
Generating functions are especially useful for manipulations with symmetric and \emph{traceless} tensors. Here, the tracelessness is enforced by further requiring that the auxiliary vectors are null:
\begin{equation}
t_{\mu_1...\mu_r}\left(x\right) \: \longrightarrow \: t\left(x,w\right) = \frac{1}{r!} t_{\mu_1...\mu_r}\left(x\right) w^{\mu_1}...w^{\mu_r}, \quad w^2=0.
\end{equation}
And for the ambient representative,
 \begin{equation}
 T_{A_1...A_r}\left(X\right) \: \longrightarrow \: T\left(X,W\right) = \frac{1}{r!} T_{A_1...A_r}\left(X\right) W^{A_1}...W^{A_r}, \quad X \cdot W = W^2=0. 
 \end{equation}
The two conditions $X \cdot W = W^2=0$ on the ambient auxiliary vector $W^A$ imply that the ambient tensor $T_{A_1...A_r}\left(X\right)$ is only determined up to radial or pure trace tensors, \textit{i.e.} it is 
an equivalence class
$T_{A_1...A_r} \: \sim \: X_{(A_1} S_{A_2...A_r)}+\eta_{(A_1A_2} S_{A_3...A_r)}$ of which one can always choose a tangential and traceless representative.

Just as the ambient representative $G_{AB}$ of the AdS metric defines a projector $\mathcal{P}^A_B$ onto tensors tangential to the hyperboloid $H_{d+1}$, one can apply  
 \begin{equation}
 \mathcal{P}_{\left\{A_1\right.}{}^{B_1}... \mathcal{P}_{A_r\left.\right\}}{}^{B_{r}} = \frac{1}{r!} \sum_{\pi}  \mathcal{P}_{A_{\pi_{1}}}{}^{B_1} ...  \mathcal{P}_{A_{\pi_{r}}}{}^{B_r} - \text{traces},
 \end{equation}
 to a general rank-$r$ ambient tensor to make it symmetric, traceless and tangential to $H_{d+1}$. The above sum is over all permutations of the indices, and the traces are subtracted using $G_{AB}$.

 When working with generating functions, we implement the above projection by a differential operator $K_A$ \cite{1926,Grigoriev:2011gp,Costa:2014kfa} 
 \begin{equation} \label{sandtcads}
 \frac{1}{r!\left(\frac{d-1}{2}\right)_{r}} K_{A_1} ...K_{A_r} W^{B_1} ...W^{B_r} = \mathcal{P}_{\left\{A_1\right.}{}^{B_1}...\mathcal{P}_{A_r\left.\right\}}{}^{B_{r}},
 \end{equation}
 where $\left(a\right)_{r} = \Gamma\left(a+r\right)/\Gamma\left(a\right)$ is the rising Pochhammer symbol. Accordingly, $K_{A}$ is tangent $\left(X^{A} K_{A} = 0\right)$, symmetric $\left(K_{A}K_{B} = K_{B} K_{A}\right)$ and traceless 
$\left( K^{A} K_{A} = 0\right)$. In this paper, $K_{A}$ is used primarily to implement contractions between symmetric and traceless tensors. Its explicit form is lengthy, and is given by \eqref{K} in appendix \ref{appendix:operators}.

By using $K_{A}$ to make symmetric and traceless contractions, involved manipulations of symmetric and traceless tensors can be made more manageable. For example, the divergence of a symmetric and traceless rank-$r$ tensor 
can be computed as\footnote{One should note that $\nabla \cdot K = K \cdot \nabla$ on the submanifold $X^2+1 = X \cdot W = W^2=0$.}
\begin{equation} \label{tldiv}
\left(\nabla \cdot T\right)\left(X,W\right) = \frac{1}{r!\left(\frac{d-3}{2}+r\right)} \left(\nabla \cdot K\right) T\left(X,W\right).
\end{equation}
Then manipulations involving the divergence of $T\left(X,W\right)$ can be simplified by combining the above with other identities involving $K_A$. A collection of those used in this work can be found in
appendix \ref{appendix:operators}.

 Throughout, we reserve $W^A$ and the corresponding intrinsic $w^{\mu}$ as auxiliary variables for tensors which are both symmetric and traceless only. For symmetric tensors, we use $U^{A}$ and $u^{\mu}$. 

\subsection{Ambient boundary tensors}
To make contact with the dual CFT in the four-point exchange computations, in this section we include the extension of the ambient formalism to fields defined on the conformal boundary of AdS$_{d+1}$.

In the boundary CFT, the operators dual to the higher-spin fields in the bulk are primary operators represented by symmetric and traceless tensors. A spin-$r$ primary field $f_{m_1 ...m_r}\left(y\right)$ of dimension
$\Delta$ is represented in the ambient formalism by a $SO\left(d+1,1\right)$-tensor $F_{A_1...A_r}\left(P\right)$ on the light cone $P^{2} = 0$ \eqref{lightcone}. This ambient tensor is also symmetric and traceless, 
as well as homogeneous of degree $-\Delta$:
\begin{equation}
F_{A_1...A_r}\left(\lambda P \right) = \lambda^{\Delta} F_{A_1...A_r}\left(P\right), \qquad \lambda > 0. 
\end{equation}
To be tangent to the light cone, $F_{A_1...A_r}\left(P\right)$ must satisfy
\begin{equation}
P^{A_1} F_{A_1...A_r}=0. \label{boundaryt}
\end{equation}
This eliminates one of the two additional components per index compared to the corresponding intrinsic field $f_{m_1 ...m_r}\left(y\right)$ in two lower dimensions. The remaining component is accounted for when one notices that anything proportional to $P^{A}$ (with the correct symmetry properties) lies in the kernel of the pull back 
\begin{equation}
f_{m_1 ...m_r}\left(y\right) =  \frac{\partial Y^{A_1}\left(y\right)}{\partial y^{m_1}} ... \frac{\partial Y^{A_r}\left(y\right)}{\partial y^{m_r}} F_{A_1...A_r}\left(P\left(y\right)\right),
\end{equation}
just like the ``pure gauge'' tensors in the ambient description of AdS space in the previous section. However, in this case being on the light cone \eqref{lightcone} means that the tangential condition \eqref{boundaryt} is not sufficient to resolve the ambiguity introduced by the non-trivial kernel. Therefore $F_{A_1...A_r}\left(P\right)$ is defined up to the addition of an arbitrary tensor proportional to $P^{A}$, and this ``gauge invariance'' takes care of the residual freedom.

Analogous to the ambient description symmetric and traceless tensors in AdS space in section \ref{sec:sandtads}, the symmetric and traceless boundary tensors can be encoded in generating polynomials
\begin{equation}
F_{A_1 ...A_r}\left(P\right) \; \longrightarrow \; F\left(P,Z\right)=\frac{1}{r!}  F_{A_1 ...A_r}\left(P\right) Z^{A_1}...Z^{A_r}, \quad Z^2=0.
\end{equation}
Tangentiality to the light cone \eqref{boundaryt} can be enforced by requiring $F\left(P,Z+\alpha P\right) = F\left(P,Z\right)$ for any $\alpha$, and the ``gauge freedom''  is represented by the constraint $Z \cdot P=0$.

Contractions between symmetric and traceless tensors can be implemented by the boundary counter part \cite{1926,Dobrev:1975ru,Grigoriev:2011gp,Costa:2011mg} of \eqref{sandtcads},
\begin{equation}
D^{A}_{Z} = \left(\tfrac{d}{2}-1+Z\cdot \frac{\partial}{\partial Z}\right)  \frac{\partial}{\partial Z_{A}} - \frac{1}{2} Z^{A} \frac{\partial^2}{\partial Z \cdot \partial Z}.
 \end{equation}

\section{Massless higher-spin bulk-to-bulk propagators}
\label{Sec:Propagators}
To compute four-point higher-spin exchanges, we first need to derive bulk-to-bulk propagators for massless bosonic higher-spin fields. To this end, in section \ref{Subsec:Fronsdal} we recapitulate free massless bosonic higher-spin fields in AdS and specify the equation that the propagators should satisfy.
We then review various features of harmonic bi-tensors, which will be useful in solving for the propagators in different gauges in the subsequent sections \ref{Sec:Propagators;Subsec:deDonder}
to \ref{Sec:Propagators;Subsec:ManifestTrace}. We conclude  by preparing the ground for the computation of the higher-spin exchange, which will be given in the form 
of a boundary conformal partial wave expansion: We briefly recall existing results for bulk-to-boundary propagators, the split representation for the bulk-to-bulk propagators and the subsequent factorisation of
exchange amplitudes. For the re-writing of the exchange amplitude, we also review the conformal partial wave expansion in conformal field theory.

\subsection{Free massless higher-spin fields in AdS}
\label{Subsec:Fronsdal}
The actions describing non-interacting spin-$s$ massless particles in the Minkowski space 
and in (A)dS were established by Fronsdal  \cite{Fronsdal:1978rb,Fronsdal:1978vb}.
In (A)dS$_{d+1}$, it reads\footnote{For the rest of the paper, when expressing tensor contractions through generating functions we implicitly set the auxiliary vector to zero -- as in \eqref{FronsdalAdS}.} 
\begin{equation}
\label{FronsdalAdS}
S=\frac{s!}{2}\int \sqrt{|g|}\,d^{d+1}x\; \varphi_{s}(x,\partial_u)\left(1-
\frac{1}4{u^2} {\partial_u\cdot \partial_u}\right){\cal F}_{s}(x,u,\nabla,\partial_u)\varphi_s(x,u)\Big|_{u=0},
\end{equation}
where ${\cal F}_{s}(x,u,\nabla,\partial_u)$ is the Fronsdal operator \cite{Metsaev:1999ui,Mikhailov:2002bp}
\begin{align} \label{Fronsdaltensor}
{\cal F}_{s}(x,u,\nabla,\partial_u)
=
\Box- m^2_s-u^2(\partial_u\cdot \partial_u)
-\;&(u\cdot \nabla)\left((\nabla\cdot\partial_u)-\frac{1}{2}(u\cdot \nabla)
(\partial_u\cdot \partial_u)
\right),
\end{align}
\begin{equation}
m_s^2\equiv
s^2+s(d-5)-2(d-2)
, \notag
\end{equation}
and $\varphi_{s}(x,u)$ is a generating function for the off-shell spin-$s$ Fronsdal field, which is a rank-$s$
symmetric double-traceless tensor 
\begin{equation} \label{gfield}
\varphi_{s}(x,u)\equiv\frac{1}{s!} \varphi_{\mu_1\mu_2\dots \mu_s}u^{\mu_1}u^{\mu_2}\dots u^{\mu_s},
\qquad
(\partial_u\cdot \partial_u)^2\varphi_{s}(x,u)
= 0.
\end{equation}
The action \eqref{FronsdalAdS} is gauge invariant with respect to transformations
\begin{equation}
\label{fronsdalgaugetr}
\delta\varphi_{s}(x,u)=(u\cdot\nabla)\varepsilon_{s-1}(x,u),
\end{equation}
where $\varepsilon_{s-1}(x,u)$ is a generating function for a rank-$\left(s-1\right)$ symmetric and traceless
gauge parameter 
\begin{equation} \label{gaugepara}
\varepsilon_{s-1}(x,u)\equiv\frac{1}{(s-1)!} \varepsilon_{\mu_1\mu_2\dots \mu_{s-1}}u^{\mu_1}u^{\mu_2}\dots u^{\mu_{s-1}},
\qquad
(\partial_u\cdot \partial_u)\varepsilon_{s-1}(x,u)
= 0.
\end{equation}
One can introduce an interaction with a conserved current by adding a term
\begin{equation}
\label{intterm}
S_{int}=-{s!}\int \sqrt{|g|}\,d^{d+1}x\; \varphi_{s}(x,\partial_u) J_{s}(x,u)\Big|_{u=0},
\end{equation}
where it is required by consistency with higher-spin gauge symmetry that the current is double-traceless and conserved up to pure trace terms. That is,
\begin{equation}
\label{currconserv}
(\partial_u\cdot \nabla)J_{s}(x,u) = u^2 Q_{s-3}(x,u)
\end{equation}
for some $Q(x,u)$. The associated equation of motion is
\begin{equation}
\label{eom}
\left(1-
\frac{1}4{u^2} {\partial_u\cdot \partial_u}\right){\cal F}_{s}(x,u,\nabla,\partial_u)
\varphi_{s}(x,u) =J_{s}(x,u).
\end{equation}

It is also possible to relax the trace constraints \eqref{gaugepara} on the gauge parameter and the corresponding double-trace constraints \eqref{gfield} on the gauge
field \cite{Francia:2002aa,Francia:2002pt,Francia:2007qt}. Consequently, the \emph{unconstrained} external current $\mathcal{J}_{s}$, with which the gauge field interacts, must be strictly conserved
\begin{equation}
\label{currconserv1}
(\partial_{u}\cdot \nabla){\cal J}_{s}(x,u)=0.
\end{equation}
This is in contrast to the double-traceless $J_{s}$ of the constrained formulation above, which are only required to be partially conserved \eqref{currconserv}. 
\subsubsection*{Bulk-to-bulk propagators}
We define the spin-$s$ bulk-to-bulk propagator $\Pi_s(x_1,u_1,x_2,u_2)$ as a bi-tensor that
determines  the spin-$s$ field sourced by the double-traceless current $J_s(x,u)$
\begin{equation*}
\varphi_s(x_1,u_1)=-s! \int \sqrt{|g|}\, d^{d+1}x_2 \;\Pi_s(x_1,u_1,x_2,\partial_{u_2})J_s(x_2,u_2)\Big|_{u_2=0}.
\end{equation*}
We see that the propagator must then satisfy
\begin{align}
\notag
\left(1-
\frac{1}4{u_1^2} {\partial_{u_1}\cdot \partial_{u_1}}\right)&{\cal F}_{s}(x_1,u_1,\nabla_1,\partial_{u_1})\Pi_{s}(x_1,u_1,x_2,u_2) =\\
\label{propdef}
& \qquad -\left\{\left\{(u_1\cdot u_2)^s\right\}\right\}\delta(x_1,x_2)
+(u_2\cdot\nabla_2)\Lambda_{s,s-1}(x_1,u_1,x_2,u_2),
\end{align}
where $\left\{\left\{\bullet\right\}\right\}$ is a projection onto $u_1$-double-traceless part:\footnote{
This makes the result also double-traceless in $u_2$ as a consequence.}
\begin{align*}
& \qquad \qquad \qquad \quad \qquad(\partial_u\cdot \partial_u)^2\left\{\left\{f(u,x)\right\}\right\}=0, \\
&\text{and} \quad \left\{\left\{f(u,x)\right\}\right\}= f(u,x)  \qquad \text{iff} \qquad (\partial_u\cdot \partial_u)^2 f(u,x)=0.
\end{align*}
In \eqref{propdef} $\Lambda_{s,s-1}$ is a bi-tensor which is traceless in tangent indices at $x_2$ and
double-traceless in tangent indices at $x_1$
\begin{align}
\notag
&(u_2\cdot \partial_{u_2})  \Lambda_{s,s-1}=s-1,\qquad (u_1\cdot \partial_{u_1})  \Lambda_{s,s-1}=s,\\
\notag \\ \notag
&\qquad (\partial_{u_2}\cdot\partial_{u_2}) \Lambda_{s,s-1}=(\partial_{u_1}\cdot\partial_{u_1})^2 \Lambda_{s,s-1}=0.
\end{align}
It acts as a pure gauge term, and will cancel out when integrating \eqref{propdef}
against any conserved current. This ambiguity leads to the ambiguity in the definition
of the propagator
\begin{equation*}
\Pi_s(x_1,u_1,x_2,u_2)\sim \Pi_s(x_1,u_1,x_2,u_2)+(u_2\cdot\nabla_2)\;{\cal E}_{2,s,s-1}(x_1,u_1,x_2,u_2),
\end{equation*}
where ${\cal E}_2$ is defined by $\Lambda$.

Fixing $\Lambda$ does not yet specify the propagator uniquely. Indeed, due to gauge
invariance \eqref{fronsdalgaugetr}, the left hand side of \eqref{propdef} is not sensitive
to  variations of the propagator of the form
\begin{equation*}
\Pi_s(x_1,u_1,x_2,u_2)\sim \Pi_s(x_1,u_1,x_2,u_2)+(u_1\cdot\nabla_1)\;{\cal E}_{1,s-1,s}(x_1,u_1,x_2,u_2),
\end{equation*}
where ${\cal E}_1$ has the same rank and trace properties as ${\cal E}_2$, but with
`1' and `2' interchanged. The latter freedom is usually fixed by imposing a gauge,
which makes the operator on the left hand side of \eqref{propdef} non-degenerate and hence,
it defines the propagator uniquely.

To summarise, the gauge ambiguity in the definition of the propagator is
\begin{align} \label{gaured}
\Pi_s(x_1,u_1,x_2,u_2) &\sim \Pi_s(x_1,u_1,x_2,u_2) \\ \notag
&+(u_1\cdot \nabla_1) {\cal E}_{1,s-1,s}(x_1,u_1,x_2,u_2)+
( u_2\cdot \nabla_2) {\cal E}_{2,s,s-1}(x_1,u_1,x_2,u_2).
\end{align}
In sections \ref{Sec:Propagators;Subsec:deDonder} to \ref{Sec:Propagators;Subsec:ManifestTrace} we solve \eqref{propdef} in different gauges, after introducing  the tools we use to do so in section
\ref{Sec:Propagators;Subsec:Split}. Note that the propagator
is symmetric only if the symmetries associated with ${\cal E}_1$ and ${\cal E}_2$
are fixed accordingly. Also, let us stress that the traceless-transverse gauge,
which can be achieved everywhere for free Fronsdal fields, cannot be achieved for the propagator 
due to presence of the source term on the right hand side of \eqref{propdef}.

\subsection{Basis of harmonic functions}
\label{Sec:Propagators;Subsec:Split}
After fixing a gauge, solving equation \eqref{propdef} for the bulk-to-bulk propagators is facilitated by working in a particular basis of AdS bi-tensorial harmonic functions,\footnote{See for example section 
4.C in \cite{Penedones:2007ns}, and more recently in a similar context: \cite{Costa:2014kfa}.} which we review in this section. The problem is then reduced to an algebraic one, \textit{i.e.} 
to identify coefficients of the basis functions in which the propagators are expanded. In this basis, we shall see that the task essentially amounts to algebraic manipulations of symmetric and traceless 
tensors, the machinery for which has already been well established (in the ambient formalism, see for example \cite{Costa:2014kfa,Grigoriev:2011gp})

The AdS bi-tensorial harmonic functions $\Omega_{\nu, \ell}\left(x_1,u_1,x_2,u_2\right)$ are spin-$\ell$ eigenfunctions of the Laplacian operator\footnote{It should be noted that the RHS of the bi-linear equation \eqref{harmoniceq} is exactly equal to zero everywhere.}
\begin{align}
\label{harmoniceq}
\left(\Box^{2}_{1}+\tfrac{\:d^2}{4}+\nu^2+\ell \right) \Omega_{\nu, \ell}\left(x_1,u_1,x_2,u_2\right) &= 0, \qquad \nu \in \mathbb{R}
\end{align}
which are also traceless and transverse
\begin{align}
\label{trlessdivfree}
\left(\partial_{u_1} \cdot \partial_{u_1}\right)\Omega_{\nu, \ell}\left(x_1,u_1,x_2,u_2\right)&=0, \qquad \left(\nabla_{1} \cdot \partial_{u_1}  \right) \Omega_{\nu, \ell}\left(x_1,u_1,x_2,u_2\right)=0.
\end{align}
Tracelessness in particular means that $\Omega_{\nu, \ell}\left(x_1,u_1,x_2,u_2\right)$ can equivalently be represented with $\Omega_{\nu, \ell}\left(x_1,w_1,x_2,w_2\right)$, $w_1^2=w^2_2=0.$

That these functions provide a basis for arbitrary spin bulk-to-bulk propagators is due to the completeness relation
\begin{align}
\left(w_1 \cdot w_2\right)^{r} \delta^{d+1}\left(x_1,x_2\right) = \sum^{r}_{\ell=0} \int^{\infty}_{-\infty} d\nu \; c_{r,\ell} \,\left(w_1 \cdot \nabla_1 \right)^{\ell} \left(w_2 \cdot \nabla_2 \right)^{\ell} \Omega_{\nu, r-\ell}\left(x_1,w_1,x_2,w_2\right),
\end{align}
where
\begin{equation}
c_{r,\ell}(\nu) = \frac{2^{\ell} \left(r-\ell+1\right)_{\ell}\left(\tfrac{d}{2}+r-\ell - \frac{1}{2}\right)_{\ell}}{\ell! \left(d+2r-2\ell -1\right)_{\ell} \left(\tfrac{d}{2}+r-\ell -i\nu\right)_{\ell}  \left(\tfrac{d}{2}+r-\ell +i\nu\right)_{\ell}}. \label{eq:complete}
\end{equation} 
Then the set of bi-tensors
\begin{align} \label{eq:basis}
 \left\{ \left(w_1 \cdot \nabla_1 \right)^{\ell} \left(w_2 \cdot \nabla_2 \right)^{\ell} \Omega_{\nu, r-\ell}\left(x_1,w_1,x_2,w_2\right) \big| \quad \nu \in \mathbb{R}, \quad \ell = 0,1,...,r \right\},
\end{align}
form a complete basis for arbitrary rank-$r$ symmetric and traceless tensors in AdS$_{d+1}$. In the following, we utilise this basis to solve for the massless higher-spin bulk-to-bulk propagators in 
particular gauges. Namely, de Donder gauge (section \ref{Sec:Propagators;Subsec:deDonder}) and a traceless gauge (section \ref{Sec:Propagators;Subsec:tracelessgauge}).

\subsection{de Donder gauge}
\label{Sec:Propagators;Subsec:deDonder}
It is often useful to eliminate gradients and divergences from the Fronsdal tensor \eqref{Fronsdaltensor}. To do so, one usually imposes  de Donder gauge
\begin{eqnarray}
\label{dedondg1}
(\nabla\cdot\partial_u)\varphi_{s}(x,u)-\frac{1}{2}(u\cdot \nabla)
(\partial_u\cdot \partial_u)\varphi_{s}(x,u)=0.
\end{eqnarray}
In this gauge, the Fronsdal tensor reads
\begin{eqnarray}
\label{FronsdaltensorDD}
{\cal F}_{s}(x,u,\nabla,\partial_u)\varphi_{s}(x,u)=(\Box- m^2_s)\varphi_{s}(x,u)-u^2(\partial_u\cdot \partial_u)\varphi_{s}(x,u).
\end{eqnarray}
The double-traceless propagator can be decomposed into two traceless parts as follows
\begin{align}
\notag
\Pi_{s}(x_1,&u_1,x_2,u_2)=\Pi^{\{0\}}_{s}(x_1,u_1,x_2,u_2)+u_1^2 u_2^2\Pi^{\{1\}}_{s-2}(x_1,u_1,x_2,u_2),\\
\label{tracesplit1} \\ \nonumber
(\partial_{u_1}\cdot\partial_{u_1})&\Pi^{\{0\}}_{s}=(\partial_{u_2}\cdot\partial_{u_2})\Pi^{\{0\}}_{s}=0=
(\partial_{u_1}\cdot\partial_{u_1})\Pi^{\{1\}}_{s-2}=(\partial_{u_2}\cdot\partial_{u_2})\Pi^{\{1\}}_{s-2}\,.
\end{align}
Analogously, we can decompose
\begin{align}
\left\{\left\{(u_1\cdot u_2)^s\right\}\right\}
\label{tracesplit2}
=\left\{(u_1\cdot u_2)^s\right\}+\frac{s(s-1)}{2(d+2s-3)}u_1^2 u_2^2\left\{(u_1\cdot u_2)^{s-2}\right\}.
\end{align}
As a consequence, \eqref{propdef} splits in two parts
\begin{align}
\label{dedonderprop1}
(\Box_1-m_s^2)\Pi^{\{0\}}_{s}&(x_1,u_1,x_2,u_2)=-\left\{(u_1\cdot u_2)^s\right\}\delta(x_1,x_2),\\ \nonumber \\ \nonumber
(\Box_1-m_t^2)\Pi^{\{1\}}_{s-2}&(x_1,u_1,x_2,u_2)=\frac{s(s-1)}{(d+2s-3)(d+2s-5)}\left\{(u_1\cdot u_2)^{s-2}\right\}\delta(x_1,x_2),
\end{align}
where
\begin{equation}
\label{tracemass}
m_t^2=s^2+(d-1)s-2.
\end{equation}
\subsubsection*{Solving for the propagator}
Both $\Pi^{\{0\}}_{s}$ and $\Pi^{\{1\}}_{s-2}$ are traceless, so we can expand them in the basis \eqref{eq:basis} 
of harmonic functions for symmetric and traceless tensors,
\begin{align}
\label{tracelesspartdeDonder}
\Pi^{\{0\}}_{s} &= \sum^{s}_{\ell=0} \int^{\infty}_{-\infty} d\nu f^{\{0\}}_{s,\ell}\left(\nu\right) \left(W_1 \cdot \nabla_1\right)^{\ell}\left(W_2 \cdot \nabla_2\right)^{\ell} \Omega_{\nu,s-\ell}\left(X_1,X_2;W_1,W_2\right),\\ \nonumber
\Pi^{\{1\}}_{s-2}& = \sum^{s-2}_{\ell=0} \int^{\infty}_{-\infty}  d\nu f^{\{1\}}_{s-2,\ell}\left(\nu\right) \left(W_1 \cdot \nabla_1\right)^{\ell}\left(W_2 \cdot \nabla_2\right)^{\ell} \Omega_{\nu,s-2-\ell}\left(X_1,X_2;W_1,W_2\right),
\end{align}
where $f^{\{0\}}_{s,\ell}$ and $f^{\{1\}}_{s-2,\ell}$ remain to be fixed by
\eqref{dedonderprop1}. Using the the equation of motion \eqref{harmoniceq} for $\Omega$, the completeness relation \eqref{eq:complete} and the commutator \eqref{i1},
it is straightforward to see that
\begin{align}
f^{\{0\}}_{s,\ell}=&\;\frac{c_{s,\ell}}{m_s^2+\tfrac{\:d^2}{4}+\nu^2+s-\ell+l(d+2s-\ell-1)},\\ \nonumber \\ 
f^{\{1\}}_{s-2,\ell}=&-\frac{s(s-1)}{(d+2s-3)(d+2s-5)}\frac{c_{s-2,\ell}}{m_t^2+\tfrac{\:d^2}{4}+\nu^2+s-\ell-2+\ell(d+2s-\ell-5)}.
\end{align}

\subsection{Traceless gauge}
\label{Sec:Propagators;Subsec:tracelessgauge}
We now derive the propagators in a traceless gauge. Since setting the trace of a higher-spin field to zero is only a partial gauge fixing, in this case we need to further restrict the form
of the traceless ansatz to obtain a unique solution to the propagator equation.

To set the trace of the spin-$s$ Fronsdal  field  to zero, the gauge parameter $\varepsilon_{s-1}\left(x,u\right)$ needs to be chosen such that 
\begin{equation}
\left(\nabla \cdot \partial_{u} \right) \varepsilon_{s-1} \left(x,u\right)  = - \left(\partial_{u} \cdot \partial_{u}\right)\varphi_{s}. 
\end{equation} 
This equation can be solved for any trace $\left(\partial_{u} \cdot \partial_{u}\right)\varphi_{s}$, after which $\left(\nabla \cdot \partial_{u} \right) \varepsilon_{s-1}$ is fixed. In this partial gauge, the field $\varphi_{s}\left(x,u\right)$ is left with residual gauge symmetry 
\begin{equation} \label{resg}
\delta\varphi_{s}(x,u)=(u\cdot\nabla)\varepsilon_{s-1}(x,u), \quad \left(\nabla \cdot \partial_{u} \right) \varepsilon_{s-1} \left(x,u\right) = \left(\partial_u \cdot \partial_u\right)  \varepsilon_{s-1} \left(x,u\right) = 0.
\end{equation}
Subject to the condition $\left(\partial_{u} \cdot \partial_{u}\right)\varphi_{s}=0$, the Fronsdal tensor then reads 
\begin{align}  \label{Tracelesseom}
& {\cal F}_{s}(x,u,\nabla,\partial_u)\varphi_{s}(x,u) \\ \nonumber
 & \qquad \qquad =\left(\Box -m^2_s\right) \varphi_{s}\left(x,u\right)- \left(u \cdot \nabla \right) \left(\nabla \cdot \partial_{u}\right) \varphi_{s} \left(x,u\right) + \frac{1}{d+2s-3} u^2 \left(\nabla \cdot \partial_{u}\right)^2 \varphi_{s}\left(x,u\right), 
\end{align}
which is simply its traceless part.

To solve \eqref{propdef} for the propagator $\Pi_{s}(x_1,u_1,x_2,u_2)$ in a traceless gauge, one needs to fully fix the residual gauge symmetry \eqref{resg}. We find that this can be done by demanding that,
in addition to the traceless condition, the propagator is symmetric under $u_1 \leftrightarrow u_2$.\footnote{Note that \eqref{propdef} does not assume this symmetry.} The equation for the propagator is then
\begin{align}
&\left(\Box_1 -m^2_s\right) \Pi_{s}(x_1,u_1,x_2,u_2) - \left(u_1 \cdot \nabla_1 \right) \left(\nabla_1 \cdot \partial_{u_1}\right) \Pi_{s}(x_1,u_1,x_2,u_2) \\ 
&\qquad \qquad \qquad \qquad +\frac{1}{d+2s-3} u^2_1 \left(\nabla_1 \cdot \partial_{u_{1}}\right)^2 \Pi_{s}(x_1,u_1,x_2,u_2)= -\left\{(u_1\cdot u_2)^s\right\}\delta(x_1,x_2), \nonumber
\end{align}
with
\begin{align}
\partial_{u_1}\cdot \partial_{u_1} \Pi_{s}(x_1,u_1,x_2,u_2) &= \partial_{u_2}\cdot \partial_{u_2} \Pi_{s}(x_1,u_1,x_2,u_2) = 0, \\ \qquad \Pi_{s}(x_1,u_1,x_2,u_2) &= \Pi_{s}(x_1,u_2,x_2,u_1).
\end{align}
In the ambient formalism, using the null auxiliary vectors $W_{1,2}$ this is
\begin{align} \label{eq:AmbientTraceless}
\left(\Box_1 -m^2_s\right)& \Pi_{s}-\frac{2}{d+2s-3} \left(W_1 \cdot \nabla_1\right)\left(\nabla_1 \cdot K_1\right) \Pi_{s}= -\left(W_1 \cdot W_2\right)^s \delta\left(X_1,X_2\right), 
\end{align}
with $\Pi_{s}\left(X_1,X_2;W_1,W_2\right) = \Pi_{s}\left(X_1,X_2;W_2,W_1\right)$.

\subsubsection*{Solving for the propagator}
Since the propagator is traceless, we can again expand in the basis \eqref{eq:basis}:
\begin{equation}
\Pi_{s} = \sum^{s}_{\ell=0} \int^{\infty}_{-\infty}  d\nu f_{s,\ell}\left(\nu\right) \left(W_1 \cdot \nabla_1\right)^{\ell}\left(W_2 \cdot \nabla_2\right)^{\ell} \Omega_{\nu,s-\ell}\left(X_1,X_2;W_1,W_2\right), \label{tlsplit}
\end{equation}
with $f_{s,\ell}\left(\nu\right)$ arbitrary functions to be determined. As in the de Donder gauge, by simply employing \eqref{harmoniceq}, \eqref{eq:complete}, \eqref{i1} and also the commutator \eqref{i2}, the propagator equation \eqref{eq:AmbientTraceless} uniquely determines these basis coefficients to be
\begin{equation}
f_{s,\ell}\left(\nu\right) = - c_{s,\ell}\left(\nu\right) \frac{d+2s-3}{\left(\ell-1\right)\left(2s+d-\ell -3\right)}\frac{1}{\nu^2+\left(s-2+\tfrac{d}{2}\right)^2}.
\end{equation}
\subsection{Manifest trace gauge}
\label{Sec:Propagators;Subsec:ManifestTrace}
In the previous two subsections, we showed how working in the basis of harmonic functions \eqref{eq:basis} allowed us to straightforwardly determine the massless higher-spin bulk-to-bulk propagators 
in de Donder gauge and the traceless gauge. If one is further able to promote the double-traceless, partially conserved currents $J_s$ to their unconstrained, strictly conserved counterparts 
$\mathcal{J}_s$,\footnote{\emph{i.e.} with $J_{s}(x,u)=\{\{{\cal J}_{s}(x,u)\}\}$. It is straightforward to show that the double-traceless part of an unconstrained, strictly conserved, current satisfies the weaker
conservation law \eqref{currconserv}.} the extra freedom recovered from the now unconstrained gauge parameters allows the gauge to be fixed further. Essentially, one is then able to remove gradient terms in the propagators.
As will be shown in section \ref{Sec:FourPointExchange;Subsec:Currents}, for the currents entering the higher-spin exchange between two pairs of scalars, this is indeed the case. In this 
section, we show how a particularly useful gauge can be reached as a consequence, which essentially eliminates the need to confront derivative operations in the exchange computation.

With the freedom to remove gradients within the propagators \eqref{tracelesspartdeDonder} and \eqref{tlsplit} already derived, it should be possible to bring them into the form
\begin{equation}
\label{eq:ManifestTrace}
\Pi_{s} = \sum^{[s/2]}_{k=0} \int^{\infty}_{-\infty}  d\nu \; g_{s,k}\left(\nu\right) \left(u_1^2\right)^{k} \left(u_2^2\right)^{k} \Omega_{\nu,s-2k},
\end{equation}
by making gauge transformations. This is the goal of this section. We refer to this as the ``manifest trace gauge'', 
because here the propagator is presented as a sum of terms, each being essentially a product of certain number of background metrics and a
harmonic function $\Omega$, which is traceless and transverse.

To reach the form \eqref{eq:ManifestTrace}, naively one might expect that, for example in \eqref{tlsplit}, one can gauge away all the terms
except the one for which $\ell=0$. However, closer inspection reveals that this is not the case: According to our conventions,
 contractions with $W$ implicitly make a projection onto the traceless part,
 so a generic term in \eqref{tlsplit} is of the form
  \begin{equation}
  \label{tomanifesttraces}
\left\{ \left(u_1 \cdot \nabla_1\right)^{\ell}\left(u_2 \cdot \nabla_2\right)^{\ell}
 \Omega_{\nu,s-\ell}(u_1,x_1,u_2,x_2)\right\}.
\end{equation}
For a general rank-$s$ tensor $T_s$, the explicit action of the operator that implements the traceless projection $\left\{\bullet\right\}$, is
 \begin{eqnarray} \label{tlprojector}
 \left\{T_{s}(x,u)\right\} = \sum_{j=0}^{[s/2]}\frac{(-1)^j}{4^j j! (\tfrac{d}{2}+s-3/2)_j}(u^2)^j (\partial_u\cdot \partial_u)^j 
 T_{s}(x,u),
 \end{eqnarray}
 which makes various contractions of its argument $T_s$. It is then clear that for non-zero $\ell$ there are terms in  \eqref{tomanifesttraces} which cannot be gauged away. For example, terms in which all $\nabla$'s are contracted because then no gradients will be present. In fact, the story is even more complicated because for other terms in the projection the commutators of derivatives that appear yield extra lower derivative
 terms, some of which are also not pure gauge. To summarise, eliminating pure gradient terms is non-trivial, and our aim in the following
 is to compute \eqref{tomanifesttraces} explicitly modulo such gradient terms. We do this by studying the details of the contractions under \eqref{tlprojector}.

It is straightforward to see that when  ${\ell}$ is odd, \eqref{tomanifesttraces} can be 
gauged away since it is a gradient. Let us then consider examples for when $\ell$ is even:
\begin{itemize}
 \item \underline{$\ell = 2$}
 
  Using \eqref{tlprojector}, the explicit form of the traceless projection for $\ell=2$ is
 \begin{align} \label{l2}
&\left\{ \left(u_1 \cdot \nabla_1\right)^{2}\left(u_2 \cdot \nabla_2\right)^{2}
 \Omega_{\nu,s-2}(u_1,u_2)\right\}&\\\nonumber
 &\quad =\left(1-\frac{u_1^2 (\partial_{u_1}\cdot \partial_{u_1})}{2(d+2s-3)}\right)
 \left(1-\frac{u_2^2 (\partial_{u_2}\cdot \partial_{u_2})}{2(d+2s-3)}\right)&
 \left(u_1 \cdot \nabla_1\right)^{2}\left(u_2 \cdot \nabla_2\right)^{2}
 \Omega_{\nu,s-2}(u_1,u_2). \nonumber
 \end{align} 
 Dropping gradient terms, it is straightforward to compute that 
\begin{equation}
\label{trace1time}
(\partial_{u_1}\cdot \partial_{u_1})(u_1\cdot \nabla_1)^2\Omega_{\nu,n}(u_1,u_2)\sim
2\left(\Box_1-n(d+n-1)\right)\Omega_{\nu,n}(u_1,u_2),
\end{equation}
where we used the traceless and divergence-less properties \eqref{trlessdivfree} of $\Omega$.\footnote{The dropping of gradient terms is denoted by $``\sim"$. Note that in appendix \ref{apptraceofcur1} we also use this notation to instead indicate that equalities hold modulo gradients \emph{and} traces.} Therefore the only terms that are not pure gradient in \eqref{l2} come from the product of the second terms in each of the brackets:
 \begin{align}
 \notag
 \frac{u_1^2 (\partial_{u_1}\cdot \partial_{u_1})}{2(d+2s-3)}(u_1\cdot \nabla_1)^2
& \frac{u_2^2 (\partial_{u_2}\cdot \partial_{u_2})}{2(d+2s-3)}(u_2\cdot \nabla_2)^2
 \Omega_{\nu,s-2}(u_1,u_2)\\
 \label{ell2almostans}
\; \sim \; & u_1^2 u_2^2\left(\frac{\Box_1-(s-2)(d+s-3)}{d+2s-3}\right)^2 \Omega_{\nu,s-2}(u_1,u_2),
\end{align}
Finally, employing the equation of motion \eqref{harmoniceq} for $\Omega$ we find
\begin{equation}
\label{ell2ans}
\left\{ \left(u_1 \cdot \nabla_1\right)^{2}\left(u_2 \cdot \nabla_2\right)^{2}
 \Omega_{\nu,s-2}(u_1,u_2)\right\} \: \sim \:
u_1^2 u_2^2 \left(\frac{\nu^2+(\tfrac{d}{2}+s-2)^2}{d+2s-3}\right)^2 \Omega_{\nu,s-2}(u_1,u_2).
\end{equation}
\item \underline{$\ell = 4$}

Explicitly, traceless projection in this case is
\begin{align}
\notag
\left\{ \left(u_1 \cdot \nabla_1\right)^{4}\left(u_2 \cdot \nabla_2\right)^{4}
 \Omega_{\nu,s-4}(u_1,u_2)\right\}& \\
 \label{justanotherformula}
 =\Big(1-\frac{u_1^2 (\partial_{u_1}\cdot \partial_{u_1})}{2(d+2s-3)}&
 +\frac{u_1^4 (\partial_{u_1}\cdot \partial_{u_1})^2}{8(d+2s-3)(d+2s-5)}\Big)
 \Big(1\leftrightarrow 2\Big)\\
 \notag
 &\qquad \quad \times \;\left(u_1 \cdot \nabla_1\right)^{4}\left(u_2 \cdot \nabla_2\right)^{4}
 \Omega_{\nu,s-4}(u_1,u_2).
 \end{align}
In order to evaluate the traces in the above, a tedious but straightforward computation shows that
\begin{align}
\notag
(\partial_{u_1}\cdot \partial_{u_1})&(u_1\cdot \nabla_1)^4\Omega_{\nu,n}(u_1,u_2)
= 2\Big( 6 (u_1\cdot \nabla_1)^2\Box_1\\
 \label{justanotherformula1}
-2(3&n^2+3dn+4d+5n+2)
(u_1\cdot \nabla_1)^2+4u^2_1\Box_1-4n(d+n-1)u^2_1\Big)\Omega_{\nu,n}(u_1,u_2),
\end{align}
and
\begin{align}
\notag
(\partial_{u_1}\cdot \partial_{u_1})^2(u_1\cdot \nabla_1)^4\Omega_{\nu,n}(u_1,u_2)
\sim & 4\Big(6\Box^2_1-4(d+3dn-n-3n^2)\Box_1\\
 \label{justanotherformula2}
&+2n(d+n-1)(2d+n+3dn+n^2)\Big)\Omega_{\nu,n}(u_1,u_2).
\end{align}
Modulo gradient terms, each bracket in 
 \eqref{justanotherformula} therefore produces a second order polynomial in $\Box$. 
 Using the equations of motion \eqref{harmoniceq}, this eventually gives
 \begin{align}
\notag
&\left\{ \left(u_1 \cdot \nabla_1\right)^{4}\left(u_2 \cdot \nabla_2\right)^{4}
 \Omega_{\nu,s-4}(u_1,u_2)\right\}\qquad \quad &\\
  \label{ell4ans}
& \qquad  \sim
\left(u_1^2\right)^2 \left(u_2^2\right)^23^2 \left(\frac{\nu^2+(\tfrac{d}{2}+s-2)^2}{d+2s-3}\right)^2 &
\left(\frac{\nu^2+(\tfrac{d}{2}+s-4)^2}{d+2s-5}\right)^2 \Omega_{\nu,s-4}(u_1,u_2).
\end{align}
\item \underline{$\ell = 2k$}

After studying explicitly cases the $\ell=2k$ with  $k=1,2$ above, we conjecture that
\begin{align}  \label{tomanifesttraces1}
\left\{ \left(u_1 \cdot \nabla_1\right)^{2k}\right.&\left.\left(u_2 \cdot \nabla_2\right)^{2k}
 \Omega_{\nu,s-2k}(u_1,x_1,u_2,x_2)\right\}\\ \notag
 &=\left(N\left(k,s\right)\right)^2(u_1^2)^k(u_2^2)^k \left(4^k \left(\frac{\tfrac{d}{2}+s-2k+i\nu}{2}\right)_k\left(\frac{\tfrac{d}{2}+s-2k-i\nu}{2}\right)_k\right)^2\\
 \notag 
 & \qquad \quad + (u_1\cdot \nabla_1) (\dots ) + (u_2\cdot \nabla_2) (\dots ),
\end{align}
where 
\begin{equation}
\label{tomanifesttraces2}
N(k,s)=\frac{(2k)!}{4^k k! (\tfrac{d}{2}+s-k-1/2)_k}
\end{equation}
and $\left(a\right)_{r} = \Gamma\left(a+r\right)/\Gamma\left(a\right)$ is the rising Pochhammer symbol.
\end{itemize}
 Let us note that in spite of the fact that equation \eqref{tomanifesttraces1} is a conjecture, the pre-factor
$N(k,s)$ is determined exactly. We explain how in the following. The combinatorial factor $N(k,s)$ can be derived by studying
only the contractions that produce the maximal power of $\Box$. The issue of non-commutativity
of covariant derivatives for this computation is irrelevant. Indeed, let us consider an analogous
computation in the flat space. Then in \eqref{trace1time} one has only a $\Box$-term
so that instead of \eqref{ell2almostans} we find
\begin{equation}
\label{ell2ansflat}
\left\{ \left(u_1 \cdot \nabla_1\right)^{2}\left(u_2 \cdot \nabla_2\right)^{2}
 \Omega_{\nu,s-2}(u_1,u_2)\right\}\Big|_{flat}\sim
u_1^2 u_2^2 \left(\frac{\Box}{d+2s-3}\right)^2 \Omega_{\nu,s-2}(u_1,u_2).
\end{equation}
Analogously, for the $\ell=4$ flat case one finds
 \begin{align}
\notag
\left\{ \left(u_1 \cdot \nabla_1\right)^{4}\left(u_2 \cdot \nabla_2\right)^{4}
 \Omega_{\nu,s-4}(u_1,u_2)\right\}\Big|_{flat}& \\
  \label{ell4ansflat}
 \sim
\left(u_1^2\right)^2 \left(u_2^2\right)^23^2& \left(\frac{\Box^2}{(d+2s-3)(d+2s-5)}\right)^2 \Omega_{\nu,s-4}(u_1,u_2).
\end{align}
This computation can be easily generalised to the case of any $\ell=2k$, and the result is
 \begin{align}
\left\{ \left(u_1 \cdot \nabla_1\right)^{2k}\left(u_2 \cdot \nabla_2\right)^{2k}
 \Omega_{\nu,s-2k}(u_1,u_2)\right\}\Big|_{flat}
  \label{ellanyansflat}
 \sim
\left(u_1^2\right)^k \left(u_2^2\right)^kN(k,s) \;\Box^{2k} \Omega_{\nu,s-2k}(u_1,u_2),
\end{align}
with $N(k,s)$ given in \eqref{tomanifesttraces2}. In fact, this justifies its explicit form.

On the other hand, in AdS the $\Box^{2k}$-term will receive lower derivative corrections,
which originate from the non-commutativity of the covariant derivatives.
By generalising the $k=1,2$ cases, what we conjecture is that these lower derivative terms are
such that after evaluating $\Box$ on $\Omega_{\nu, s-2k}$ one finds
\begin{equation*}
\left((\nu^2+(\tfrac{d}{2}+s-2)^2)(\nu^2+(\tfrac{d}{2}+s-4)^2) \dots (\nu^2+(\tfrac{d}{2}+s-2k)^2)\right)^2.
\end{equation*}
Combining this with the pre-factor found previously, one obtains \eqref{tomanifesttraces1}.

Given \eqref{tomanifesttraces1}, by using the freedom from using unconstrained conserved currents to drop gradients we can rewrite the de Donder \eqref{tracelesspartdeDonder} and traceless \eqref{tlsplit} propagators in a gauge with manifest trace structure. The two results coincide, and the coefficients $g_{s,k}\left(\nu\right)$ in \eqref{eq:ManifestTrace} are
\begin{align}
\notag
g_{s,0}&=\frac{1}{(\tfrac{d}{2}+s-2)^2+\nu^2},\\
\label{answertrgauge}
g_{s,k}&=-\frac{\left(1/2\right)_{k-1}}{2^{2k+3}\cdot k!}\frac{(s-2k+1)_{2k}}{(\tfrac{d}{2}+s-2k)_k (\tfrac{d}{2}+s-k-3/2)_k}\\
\notag
&\qquad\times \;
\frac{\left({(\tfrac{d}{2}+s-2k+i\nu)}/{2}\right)_{k-1}\left({(\tfrac{d}{2}+s-2k-i\nu)}/{2}\right)_{k-1}}{\left({(\tfrac{d}{2}+s-2k+1+i\nu)}/{2}\right)_{k}\left({(\tfrac{d}{2}+s-2k+1-i\nu)}/{2}\right)_{k}}, \quad k\ne 0.
\end{align}
The fact that using the conjectured formula  \eqref{tomanifesttraces1} gives agreement
between two propagators computed independently in different gauges serves as an additional
argument in favour of \eqref{tomanifesttraces1}.

\subsection{Bulk-to-boundary propagators}
\label{sec:3pt}
For the external legs of the Witten diagrams, we need the form of the bulk-to-boundary propagators for scalars \cite{Witten:1998qj}.  In light of the exchange amplitude factorisation (see for example figure \ref{fig:factor}), 
we will also need bulk-to-boundary propagators for higher-spin fields \cite{Mikhailov:2002bp}. In the ambient formalism, the bulk-to-boundary propagator for a symmetric spin-$s$ field of dimension $\Delta$ is 
\begin{equation}
\Pi_{\Delta,s}\left(X,P;U,Z\right) = \mathcal{C}_{\Delta,s} \frac{\left(2 \left(Z \cdot X\right)\left(P \cdot U\right) +\left(U \cdot Z\right)\left(-2P \cdot X\right)\right)^{s}}{\left(-2 X \cdot P\right)^{\Delta+s}} \label{btobound}
\end{equation}
where the normalisation constant 
\begin{equation}
\mathcal{C}_{\Delta,s}  = \frac{\left(s+\Delta-1\right)\Gamma\left(\Delta\right)}{2 \pi^{d/2} \left(\Delta-1\right) \Gamma\left(\Delta+1-\tfrac{d}{2}\right)}, 
\end{equation}
is fixed by consistency with the corresponding bulk-to-bulk propagator. 

 A useful observation which we employ for the evaluation of the bulk amplitudes, is that bulk-to-boundary propagators of non-zero spin can be obtained with the knowledge of the scalar propagator with the same dimension. Indeed,
\begin{align} \label{bbiter}
\Pi_{\Delta,s}\left(X,P;U,Z\right) &= \frac{\mathcal{C}_{\Delta,s}}{\left(\Delta\right)_{s}} \left(\mathcal{D}_{P}\right)^{s} \frac{1}{\left(-2 X \cdot P\right)^{\Delta}} \\ \nonumber
& = \frac{\left(s+\Delta-1\right)}{\left(\Delta-1\right)\left(\Delta\right)_{s}} \left(\mathcal{D}_{P}\right)^{s} \Pi_{\Delta,0}\left(X,P\right),
\end{align}
where the differential operator $\mathcal{D}_{P}$ is given by \cite{Costa:2011dw}
\begin{equation} \label{DP}
\mathcal{D}_{P} = \left(Z \cdot U\right) \left(Z \cdot \frac{\partial}{\partial Z} - P \cdot \frac{\partial}{\partial P}\right) + \left(P \cdot U\right) \left( Z\cdot \frac{\partial}{\partial P}\right).
\end{equation}
All the above carries over to symmetric and traceless propagators, for which one simply replaces $U \rightarrow W$. 
 \subsection{Split representation of bulk-to-bulk propagators}
 \label{subsec:split}
 In this subsection we introduce the \emph{split representation} of bulk-to-bulk propagators, in which they are expressed as a product of bulk-to-boundary propagators \eqref{btobound} integrated over a common boundary point. As will become clear, this representation plays a crucial role in the computation of the four-point exchange. Although this can be motivated group-theoretically \cite{Dobrev:1998md,Leonhardt:2003qu,Leonhardt:2003sn}, it can be viewed as a consequence of a similar re-writing of the harmonic functions $\Omega_{\nu,\ell}$ in which the propagators are expanded:\footnote{For a similar application of harmonic functions, see for example \cite{Costa:2014kfa} and also references within \cite{Penedones:2007ns}.}
\begin{align}
&\Omega_{\nu,\ell}\left(X_1,X_2;W_1,W_2\right) \label{split} \\ 
&\hspace*{2.2cm} = \frac{\nu^2}{\pi \ell! \left(\tfrac{d}{2}-1\right)_{\ell}} \int_{\partial \text{AdS}} dP \; \Pi_{d/2+i\nu,\ell}\left(X_1,P;W_1,D_{Z}\right)\Pi_{d/2-i\nu,\ell}\left(X_2,P;W_2,Z\right) \nonumber
\end{align}
which is an integral of a product of spin-$\ell$ bulk-to-boundary propagators \eqref{btobound} of complex dimensions $\tfrac{d}{2} \pm i\nu$. When this form of the harmonic functions is inserted into the expressions for the bulk-to-bulk propagators, we refer to this as their \emph{split representation}. For the traceless gauge, this is
\begin{align} \label{splittl}
&\Pi_{s} = \sum^{s}_{\ell=0} \int^{\infty}_{-\infty}  d\nu f_{s,\ell}\left(\nu\right) \frac{\nu^2}{\pi \left(s-\ell\right)! \left(\tfrac{d}{2}-1\right)_{s-\ell}} \\\nonumber
& \times \int_{\partial \text{AdS}}dP  \left(W_1 \cdot \nabla_1\right)^{\ell} \; \Pi_{d/2+i\nu,s-\ell}\left(X_1,P;W_1,D_{Z}\right) \left(W_2 \cdot \nabla_2\right)^{\ell} \Pi_{d/2-i\nu,s-\ell}\left(X_2,P;W_2,Z\right),
\end{align}
and for the manifest trace gauge
\begin{align} \label{splitmt}
&\Pi_{s} = \sum^{[s/2]}_{k=0} \left(U_1^2\right)^{k} \left(U_2^2\right)^{k} \int^{\infty}_{-\infty}  d\nu \;g_{s,k}\left(\nu\right)  \frac{\nu^2}{\pi \left(s-2k\right)! \left(\tfrac{d}{2}-1\right)_{s-2k}} \\ \nonumber
&\qquad \qquad  \qquad\times \int_{\partial \text{AdS}} dP \; \left\{ \Pi_{d/2+i\nu,s-2k}\left(X_1,P;U_1,D_{Z}\right)\right\} \left\{ \Pi_{d/2-i\nu,s-2k}\left(X_2,P;U_2,Z\right)\right\},
\end{align}
where the bulk-to-boundary propagators above are traceless, as indicated by the braces for the bulk indices. It is in these two gauges that we compute the four-point exchange in section \ref{Sec:FourPointExchange}.

\begin{figure}[t]
 \centering
\includegraphics[width=0.25\linewidth]{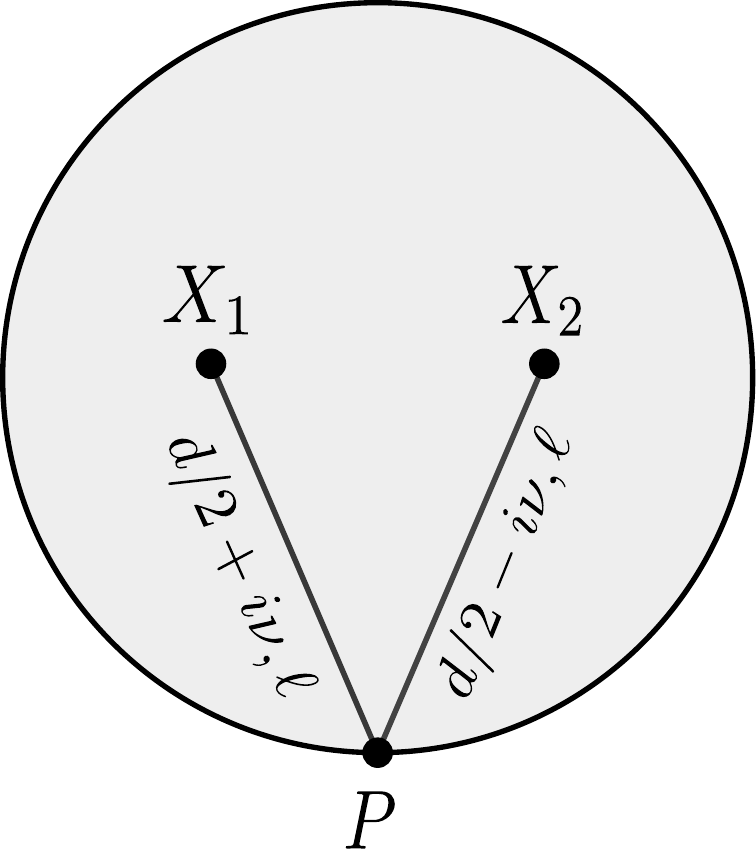} 
 \caption{Split representation of AdS harmonic function $\Omega_{\nu,\ell}$ as a product of two spin-$\ell$ bulk-to-boundary propagators with dimensions $\tfrac{d}{2}\pm i\nu$. The boundary point $P$ is integrated over.}
\label{fig:split}
\end{figure}

  This representation is depicted schematically in figure \ref{fig:split}.

\subsubsection*{Factorisation of exchange amplitudes}
The significance of the split representation of the bulk-to-bulk propagators introduced above becomes clear once one studies its impact on the form of the four-point exchange diagram. The general form of the spin-$s$ exchange is 
\begin{align} \label{genexch}
& \mathcal{A}_{s, \phi}\left(P_1,P_2 ;P_3,P_4\right) \\ 
& \: = \left(g_{\phi \phi s}\right)^2 \int_{\text{AdS}} dX_1\int_{\text{AdS}} dX_2 \; \Pi_{s}\left(X_1,\partial_{U_1} ;X_2,\partial_{U_2}\right) {\cal J}_{s}\left(X_1,U_1 ;P_{1},P_{2}\right) {\cal J}_{s}\left(X_2,U_2 ;P_{3},P_{4}\right), \nonumber
\end{align}
where ${\cal J}_{s}$ is an unconstrained spin-$s$ conserved current bilinear in the real scalar $\phi$, the precise form of which we introduce section \ref{Sec:FourPointExchange;Subsec:Currents}. Being a sum of integral products of two bulk-to-boundary propagators, it is therefore clear that the use of the split representations \eqref{splittl} and \eqref{splitmt} for the bulk-to-bulk propagator leads to the decomposition of the exchange into products of 
two three-point amplitudes involving the scalar interacting with a field of spin --- for which methods to compute are already known. This is illustrated schematically in figure \ref{fig:factor}. 
The decomposition of the four-point exchange in this manner is directly analogous
to the conformal partial wave expansion of correlation functions in conformal field theory, in which four-point correlation functions decompose in the same way into products of three-point functions.
This therefore lays the ground for an eventual comparison with the corresponding CFT result, once all contributing bulk diagrams are included (figure \ref{fig:total4pt}). In view of bringing the bulk
computations directly into this form, in 
this section we briefly recall existing results for these three-point Witten diagrams in the ambient formalism, and review the conformal partial wave expansion in conformal field theory. 

\subsubsection*{Three-point Witten diagrams}
It is enough to review the three-point Witten diagrams of two real scalars and a higher-spin field interacting with an \emph{elementary} cubic vertex,
\begin{align}
g_{\phi_1 \phi_2 \ell} \; \ell! \int_{\text{AdS}} \sqrt{g}\; d^{d+1}x\; \varphi_{\ell} \left(x,\partial_u\right) \phi_1 \left(x\right) \left(u \cdot \nabla\right)^{\ell} \phi_2 \left(x\right). \label{simpvertex}
\end{align}
The three-point bulk integrals encountered in the decomposition under the split representation can be expressed in combinations of these building blocks. Moreover, since the higher-spin field appearing in the vertex will be traceless, it is sufficient to recall the traceless contribution to these amplitudes. I.e. we take $u \rightarrow w$. Precisely this contribution to such three-point Witten diagrams was recently computed in the ambient formalism in \cite{Costa:2014kfa}.\footnote{Note that in \cite{Costa:2014kfa} and similar works, the shorthand $d = 2h$ is often employed.} For an elementary interaction \eqref{simpvertex} between two scalars $\phi_{1,2}\left(x\right)$ and a traceless spin-$\ell$ field $\varphi_{\ell}\left(x\right)$ with dimensions $\Delta_{1,2}$ and $\Delta_\ell$ respectively, these are\footnote{Here, the CFT operators are normalised such that they have unit two-point function
\begin{equation} \langle \mathcal{O}_{\Delta,\ell}\left(P_1,Z_1\right) \mathcal{O}_{\Delta,\ell}\left(P_2,Z_2\right) \rangle = \frac{\left(\left(-2 P_1 \cdot P_2\right)\left(Z_2 \cdot Z_1\right)+2\left(Z_2 \cdot P_1\right)\left(Z_1 \cdot P_2\right)\right)^{\ell}}{\left(-2 P_1 \cdot P_2\right)^{\Delta+\ell}}.
\end{equation}}
\begin{align}\label{3pt}
&\frac{g_{\phi_1 \phi_2 \ell}}{\sqrt{\mathcal{C}_{\Delta_1}\mathcal{C}_{\Delta_2}\mathcal{C}_{\Delta_\ell,\ell}}} \int_{\text{AdS}} dX \; \Pi_{\Delta,\ell}\left(X,P_3;K,Z\right)\Pi_{\Delta_1,0}\left(X,P_1\right) \left(W \cdot \nabla\right)^\ell\Pi_{\Delta_2,0}\left(X,P_2\right) \\ \nonumber
& \qquad \qquad \qquad= \frac{g_{\phi_1 \phi_2 \ell}}{\sqrt{\mathcal{C}_{\Delta_1}\mathcal{C}_{\Delta_2}\mathcal{C}_{\Delta_\ell,\ell}}} b\left(\Delta_1,\Delta_2,\Delta_\ell,\ell\right) \frac{\left(\left(Z \cdot P_1\right)P_{23}-\left(Z \cdot P_2\right)P_{13}\right)^{\ell}}{P_{12}^{\frac{\Delta_1+\Delta_2-\Delta_\ell+\ell}{2}}P_{13}^{\frac{\Delta_1-\Delta_2+\Delta_\ell+\ell}{2}}P_{23}^{\frac{\Delta_2-\Delta_1+\Delta_\ell+\ell}{2}}},
\end{align}
where
\begin{align} \label{bpw}
& b\left(\Delta_1,\Delta_2,\Delta_\ell,\ell\right) \\ 
& = \mathcal{C}_{\Delta_1}\mathcal{C}_{\Delta_2}\mathcal{C}_{\Delta_\ell,\ell} \frac{\pi^{d/2}\Gamma\left(\frac{\Delta_1+\Delta_2+\Delta_\ell-d+\ell}{2}\right)\Gamma\left(\frac{\Delta_1+\Delta_2-\Delta_\ell+\ell}{2}\right)\Gamma\left(\frac{\Delta_1+\Delta_\ell-\Delta_2+\ell}{2}\right)\Gamma\left(\frac{\Delta_2+\Delta_\ell-\Delta_1+\ell}{2}\right)}{2^{1-\ell}\Gamma\left(\Delta_1\right)\Gamma\left(\Delta_2\right)\Gamma\left(\Delta_\ell+\ell\right)}, \nonumber
\end{align}
and we introduced the shorthand $P_{ij} = -2P_{i} \cdot P_{j}$. 

For a given triplet of interacting fields in AdS, regardless of the form of the bulk cubic interaction vertex, the tensorial structure of the corresponding tree-level three-point Witten diagram is uniquely fixed by the boundary conformal symmetry. The task is then to determine the overall coefficient, which is vertex-dependent. The cubic interactions used in this paper for the four-point exchange are more complicated than \eqref{simpvertex} used in the above. This is because they are current interactions \eqref{intterm}, and to ensure current conservation their explicit form is more involved. However, as mentioned above, for the three-point decomposition of the exchange it is possible to express the coefficient of these amplitudes in terms of those for the elementary vertices.
\subsubsection*{Conformal partial wave expansion}
In conformal field theory, the conformal partial wave expansion (CPWE) of a four-point function of scalar primary operators $\mathcal{O}_{\Delta_i}\left(P\right)$ of conformal dimensions $\Delta_i$ is the decomposition \cite{Dobrev:1975ru}
\begin{equation} \label{cpwe}
\langle \mathcal{O}_{\Delta_1}\left(P_1\right) \mathcal{O}_{\Delta_2}\left(P_2\right) \mathcal{O}_{\Delta_3}\left(P_3\right) \mathcal{O}_{\Delta_4}\left(P_4\right) \rangle = \sum^{\infty}_{\ell=0} \int^{\infty}_{-\infty} d\nu\; b_{\ell}\left(\nu\right)\; F_{\nu,\ell}\left(u,v\right),
\end{equation}
where $F_{\nu,\ell}\left(u,v\right)$ is a \emph{conformal partial wave}. This can be written as a product of two three-point functions integrated over a common boundary point 
\begin{align} \label{partialwave}
&F_{\nu,\ell}\left(u,v\right) \\  \nonumber
& = \frac{1}{\beta_{\nu,\Delta_i,\ell}} \int_{\partial \text{AdS}} dP_{5} \langle \mathcal{O}_{\Delta_1}\left(P_1\right)\mathcal{O}_{\Delta_2}\left(P_2\right)\mathcal{O}_{d/2+i\nu,\ell}\left(P_5,D_Z\right)\rangle \langle \mathcal{O}_{d/2-i\nu,\ell}\left(P_5,Z\right) \mathcal{O}_{\Delta_3}\left(P_3\right) \mathcal{O}_{\Delta_4}\left(P_4\right)\rangle.
\end{align}
and is a function of the cross ratios
\begin{equation}
u = \frac{P_{12}P_{34}}{P_{13}P_{24}}, \qquad v = \frac{P_{14}P_{23}}{P_{13}P_{24}}.
\end{equation}
In this definition of the CPWE we adopt the conventions of \cite{Costa:2014kfa}, where it is also presented in the ambient formalism. In particular, we take on their normalisation
$\beta_{\nu,\Delta_i,\ell}$\footnote{The explicit form of which is given in the appendix, equation \eqref{beta_costa}.} of the conformal partial wave \eqref{partialwave}, for which the three 
point functions have unit coefficient
\begin{align} \label{canon3pt}
\langle \mathcal{O}_{\Delta_1}\left(P_1\right)\mathcal{O}_{\Delta_2}\left(P_2\right) \mathcal{O}_{d/2 \pm i\nu,\ell}\left(P_5,Z\right)\rangle = \frac{\left(\left(Z \cdot P_1\right)P_{25}-\left(Z \cdot P_2\right)P_{15}\right)^{\ell}}{P_{12}^{\frac{\Delta_1+\Delta_2+d/2\mp i\nu+\ell}{2}}P_{15}^{\frac{\Delta_1-\Delta_2+d/2 \pm i\nu+\ell}{2}}P_{25}^{\frac{\Delta_2-\Delta_1+d/2 \pm i\nu+\ell}{2}}}.
\end{align}
It is our goal in the computation of the four-point exchange to bring the decomposition of the exchange amplitude under the split representation precisely into this form of the CPWE. Since
the three-point functions in \eqref{cpwe} are normalised with unit coefficient, it is essentially the task of finding the overall coefficients of the three-point amplitudes appearing in the decomposition,
which are subsequently absorbed into $b_{\ell}\left(\nu\right)$. This is facilitated by expressing the amplitudes in terms of those for the elementary vertices introduced in the previous section.
\section{Four-point exchange}
\label{Sec:FourPointExchange}
\begin{figure}[h]
 \centering
\includegraphics[width=0.35\linewidth]{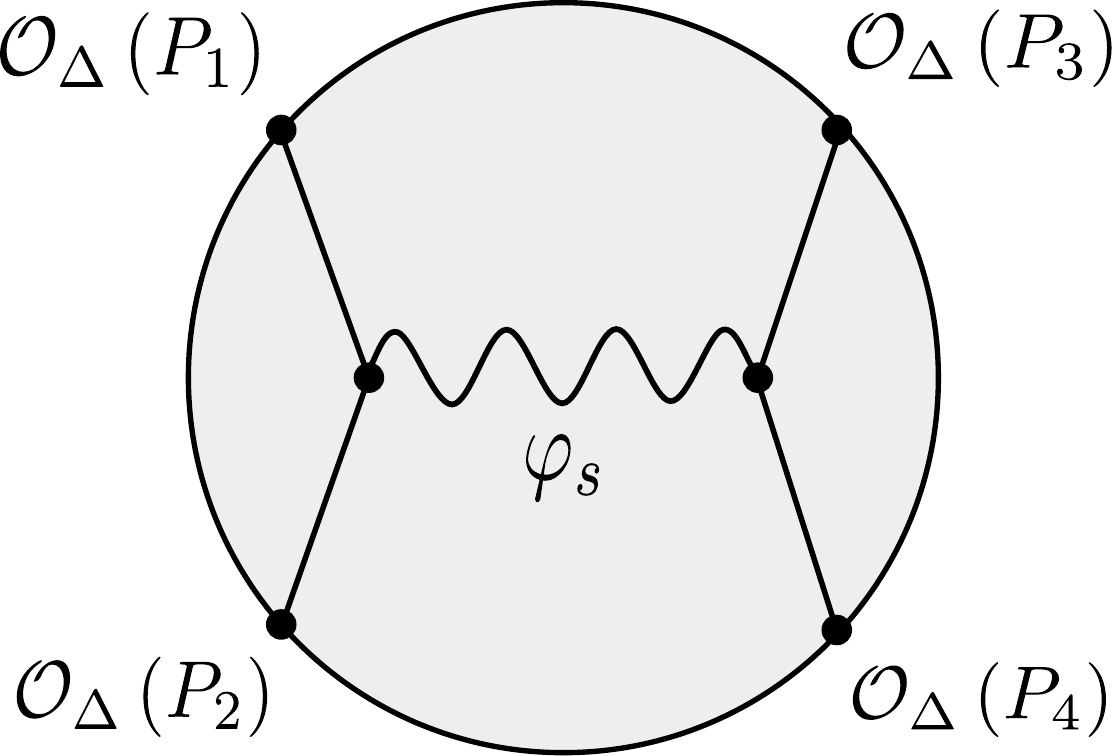} 
 \caption{Four-point exchange in the ``s-channel'' of a spin-$s$ gauge boson between two pairs of real scalars $\phi$, which have dual CFT operator $\mathcal{O}_{\Delta}$.}
\label{fig:exchange_s}
\end{figure}

 In this section we combine the results for the propagators in the previous to compute the exchange of a single spin-$s$ gauge boson between pairs of real scalars $\phi\left(X\right)$ in AdS$_{d+1}$, illustrated in figure \ref{fig:exchange_s} for the ``s-channel''. As emphasised in section \ref{subsec:split}, we use the split representation for the spin-$s$ bulk-to-bulk propagator, which allows to express the resultant amplitude in the form of a conformal partial wave expansion \eqref{cpwe} on the boundary.

 Before we proceed, the precise form of current appearing in the cubic interaction \eqref{intterm} of the exchange must be established. This is carried out in the next section. In the subsequent sections, we compute the exchange amplitude in the traceless gauge (section \ref{Sec:FourPointExchange;Subsec:TracelessGauge}) and the manifest trace gauge (section \ref{Sec:FourPointExchange;Subsec:ManifestTraceGauge}). In section \ref{subsec:checks}, we verify for explicit examples that these two gauges give the same results. In section \ref{subsec:improvements}, we comment on the effect of improvements to the currents on the exchange amplitude.
\subsection{Currents}
\label{Sec:FourPointExchange;Subsec:Currents}

Before computing the four-point higher-spin exchange, the cubic vertex between the pair of scalars and the spin-$s$ gauge field has to be specified. 
This is the current interaction \eqref{intterm}, and in this section we recall how the explicit form of the currents can be obtained: We briefly review the construction of \cite{Bekaert:2010hk},
where it was shown how the unconstrained covariantly conserved currents ${\cal J}_{s}$
in AdS${}_{d+1}$ can be built. This construction strongly relies on ambient approach
to AdS${}_{d+1}$ and demonstrates its power: The currents are obtained by projecting
the well-known Berends, Burgers and van Dam conserved currents in flat space \cite{Berends:1985xx} onto $H_{d+1}$,\footnote{Various explicit sets of (conformal) conserved currents were provided in \cite{Anselmi:1999bb,Konstein:2000bi,Gelfond:2006be,Manvelyan:2009tf,Fotopoulos:2007yq}.}
which completely avoids any difficulties
related to the non-commutativity of derivatives in curved space.
Then the double-traceless part of these currents, $J_{s}=\{\{{\cal J}_{s}\}\}$,
provides the current that satisfies \eqref{currconserv}. 
\subsubsection*{Scalar fields in AdS}
Since the relevant currents are bilinears in the bulk scalar, we first recall the representation of scalar fields in AdS and their ambient formulation.

The lowest weight unitary irreducible scalar representations of $so(d,2)$ 
\begin{equation}
\label{massdimension}
\Box\phi(x)-m^2\phi(x)=0, \qquad m^2\equiv \Delta (\Delta-d)
\end{equation}
can be realised as
the evaluation $\phi(x)$ on AdS${}_{d+1}$ of
ambient homogeneous harmonic functions 
$\Phi(X)$
\begin{equation}
\label{masslessamb}
(\partial_X\cdot \partial_X) \Phi(X) =0, \qquad      \Phi(X)=\left(X^2\right)^{-\frac{\Delta}{2}}\phi(x).
\end{equation} 
For later purpose, let us denote 
\begin{equation}
\Phi^\dagger(X)= \left(X^2\right)^{-\frac{\Delta_-}{2}}\phi(x), \label{dagger}
\end{equation}
where $\Delta_-=d-\Delta$ and we assume $\Delta\ge\Delta_-$. Throughout, $\Delta$ will often be referred to as $\Delta_+$.
\subsubsection*{The currents}
It is then straightforward to check that for any 
ambient massive scalar fields $\Phi_1(X)$ and $\Phi_2(X)$ of the same mass $M$
\begin{equation*}
(\partial_X\cdot \partial_X-M^2) \Phi_1(X)=0=(\partial_X\cdot \partial_X-M^2) \Phi_2(X)
\end{equation*}
the currents, given by the generating function \cite{Bekaert:2009ud}
\begin{equation}
\label{ambcur1}
I(X,U)=\Phi_1(X+U)\Phi_2(X-U)
\end{equation}
are conserved with respect to flat ambient space derivative
\begin{equation}
(\partial_X\cdot \partial_U)I(X,U)= 0.
\end{equation}
Explicitly, the rank-$s$ current generating function is given by
\begin{equation}
\label{rankscur1}
I_s(X,U)=\sum_{k=0}^s \frac{(-1)^k}{k!(s-k)!}
(U\cdot \partial_X)^{s-k}\Phi_1(X)(U\cdot \partial_X)^k\Phi_2(X),
\end{equation}
which is obtained by extracting from \eqref{ambcur1} the $\mathcal{O}\left(U^s\right)$ coefficient. 

However, a conserved current in the flat ambient space does not necessarily define a conserved current in the theory on AdS. In other words, 
the pull-back of \eqref{ambcur1} onto the AdS manifold is not in general conserved with respect to the AdS covariant derivative -- it must satisfy some 
additional constraints, which we specify in the following.

We require \eqref{ambcur1} to be conserved with respect to the ambient representative \eqref{cpvdercur} of the AdS covariant derivative, $\nabla = {\cal P} \circ \partial \circ {\cal P}$. First, the ambient current projection
is of the form
\begin{equation}
{\cal P} \; I(X,U) = I(X,U)+ (X\cdot U) L(X,U)
\end{equation}
for some $L(X,U)$.
Further, the commutation relation 
\begin{equation*}
[\left(\partial_X\cdot\partial_U),X\cdot U\right]=X\cdot\partial_X+U\cdot\partial_U+d+2
\end{equation*}
implies that 
\begin{align}
\label{conservpr}
(\partial_X\cdot \partial_U){\cal P} \; I(X,&U)
= (X\cdot U)\left(\partial_X\cdot\partial_U\right)
L(X,U)+
\left(X\cdot\partial_X+U\cdot\partial_U+d+2\right) L(X,U).
\end{align}
The first term in \eqref{conservpr} is transversal to $H_{d+1}$, so it drops out upon the second projection in the covariant derivative \eqref{cpvdercur}. We then demand that $I\left(X,U\right)$ is
conserved in AdS,
\begin{equation}
 \left( \nabla \cdot \partial_{U} \right) I\left(X,U\right) = 0,
\end{equation}
which yields the condition:
\begin{equation*}
\left(X\cdot\partial_X+U\cdot\partial_U+d+2\right) L(X,U)=0.
\end{equation*}
This can be satisfied by imposing the following homogeneity condition on the current
\begin{equation}
\label{homcur}
\left(X\cdot\partial_X+U\cdot\partial_U+d\right) I(X,U)=0.
\end{equation}
We have just shown that the current \eqref{ambcur1}
is covariantly conserved if it obeys \eqref{homcur}. In particular for $\Phi_1=\Phi$, and $\Phi_2=\Phi^{\dagger}$ as introduced in \eqref{dagger}, the current
\begin{equation}
\label{ambcur2}
{\cal J}(X,U)=\Phi(X+U)\Phi^\dagger(X-U)
\end{equation}
is covariantly conserved. This can also be rewritten in intrinsic AdS terms \cite{Bekaert:2010hk}, which requires the expression of 
ambient partial derivatives in terms of covariant derivatives of AdS and the metric. For the following computations of the
exchange, it is enough to know that it is of the form
\begin{equation}
\label{rankscur2}
{\cal J}_s(x,u)=\sum_{k=0}^s \frac{(-1)^k }{k!(s-k)!}
(u\cdot \nabla)^{s-k}\phi(x)(u\cdot \nabla)^k\phi(x)+ u^2 \left(\dots\right).
\end{equation}

For the manifest trace gauge (section \ref{Sec:FourPointExchange;Subsec:ManifestTraceGauge}), an important observation is that multiple traces of the currents \eqref{ambcur2} 
can be expressed in terms of currents themselves:
\begin{equation}
\left(\partial_u\cdot \partial_u\right)^k{\cal J}_s(x,u)
\label{repeatedtr}
= \tau_{s,k}\left(\nu\right){\cal J}_{s-2k}(x,u) + (u\cdot\nabla)(\dots) +  u^2  \cdot (\dots),
\end{equation}
where
\begin{align}
\notag
\tau_{s,k}\left(\nu\right)=\sum_{m=0}^k 
\frac{2^{2k}\cdot k!}{m!(k-m)!}
(\Delta-\tfrac{d}{2}-k+m&+1/2)_{k-m}\\
\label{beta}
\times\;\left(\frac{\tfrac{d}{2}+s-2m+1+i\nu}{2}\right)_{m}&\left(\frac{\tfrac{d}{2}+s-2m+1-i\nu}{2}\right)_{m},
\end{align}
 and $\left(a\right)_{r} = \Gamma\left(a+r\right)/\Gamma\left(a\right)$ is the rising Pochhammer symbol. We do not specify the gradient and trace terms 
 \begin{equation}
\label{equivalence1}
(u\cdot\nabla)(\dots) \quad \text{and} \quad  u^2  \cdot (\dots),
\end{equation}
as they will not play a role in the exchange. We derive \eqref{repeatedtr} in 
 appendices \ref{apptraceofcur} and  \ref{apptraceofcur1}.

\subsection{Traceless gauge}
\label{Sec:FourPointExchange;Subsec:TracelessGauge}
Now that all relevant ingredients for the computation of the exchange amplitude have been established, we compute the exchange in the traceless gauge. In the next section this will also be carried out in the manifest trace gauge.

For the spin-$s$ bulk-to-bulk propagators in the traceless gauge \eqref{tlsplit}, the exchange amplitude \eqref{genexch} takes the form
\begin{align} \label{exchtr}
&\mathcal{A}_{s,\phi}\left(P_1,P_2 ; P_3,P_4\right) \\ \nonumber
& = \left(\frac{g_{\phi \phi s}}{s!\left(\tfrac{d}{2}-\frac{1}{2}\right)_{s}}\right)^2 \sum^{s}_{\ell=0} \int^{\infty}_{-\infty} d\nu f_{s,\ell}\left(\nu\right) \int_{\text{AdS}} dX_1 \int_{\text{AdS}} dX_2 \; \mathcal{J}_{s}\left(X_1,K_1;P_{1},P_{2}\right) \mathcal{J}_{s}\left(X_2,K_2;P_{3},P_{4}\right) \\ \nonumber
& \hspace{6cm} \times \left(W_1\cdot \nabla_1\right)^{\ell} \left(W_2 \cdot \nabla_2 \right)^{\ell} \Omega_{\nu,s-\ell}\left(X_1,W_1;X_2,W_2\right), \\ \nonumber
\end{align}
where here we make use of the operator $K_{1,2}$ for the contraction with the symmetric and traceless propagator, hence the prefactor. Note that only the first term of the currents \eqref{rankscur2} contribute, since the traceful $u^2\left(...\right)$ terms drop out upon contraction with the traceless bulk-to-bulk propagator.

Employing the split representation \eqref{splittl} of the bulk-to-bulk propagator, the amplitude then reduces to a sum of products of two disjoint bulk integrals
\begin{align}
  \sum^{s}_{\ell=0} & \int^{\infty}_{-\infty} d\nu f_{s,\ell}\left(\nu\right) \frac{\nu^2}{\pi \left(s-\ell \right)!\left(\tfrac{d}{2}-1\right)_{s-\ell}} \int_{\partial \text{AdS}} dP_5  \\ \nonumber
&\qquad \times \frac{g_{\phi \phi s}}{s!\left(\tfrac{d}{2}-\frac{1}{2}\right)_{s}} \int_{\text{AdS}} dX_1 \; \mathcal{J}_{s}\left(X_1,K_1;P_{1},P_{2}\right) \left(W_1 \cdot \nabla_1\right)^{\ell} \Pi_{d/2+i\nu,s-\ell}\left(X_1,P_5;W_1,D_Z\right) \\ \nonumber
&\qquad \times \frac{g_{\phi \phi s}}{s!\left(\tfrac{d}{2}-\frac{1}{2}\right)_{s}} \int_{\text{AdS}} dX_2 \; \mathcal{J}_{s}\left(X_2,K_2;P_{3},P_{4}\right) \left(W_2 \cdot \nabla_2\right)^{\ell} \Pi_{d/2-i\nu,s-\ell}\left(X_2,P_5;W_2,Z\right),
\end{align}
integrated over the common boundary point $P_5$. In light of this decomposition in terms of three-point bulk integrals, our goal is to express it as a CPWE \eqref{cpwe} on the boundary. For a given term in the decomposition above, considering only the bulk integrals it can be seen by symmetry that they will each be proportional to the appropriate $\langle \mathcal{O}_{\Delta}\left(P_i\right) \mathcal{O}_{\Delta}\left(P_j\right) \mathcal{O}_{d/2\pm i\nu,s-\ell}\left(P,Z\right)\rangle$ upon their evaluation, consistent with the CPWE. What is not immediate is the overall coefficient. We extract it by simply computing the integrals, which is facilitated by the methods we reviewed in section \ref{subsec:split}, and also operations with symmetric and traceless tensors (appendix \ref{appendix:operators}). Relegating the details to appendix \ref{app:tracelessexch}, we find that
\begin{align} \label{tlpwe}
&\hspace{3cm}\mathcal{A}_{s,\phi}\left(P_1,P_2; P_3,P_4\right) =  \sum^{s}_{\ell=0} \int^{\infty}_{-\infty} d\nu \; b_{s-\ell}\left(\nu\right) F_{\nu,s-\ell}\left(u,v\right),    \\ \nonumber 
&F_{\nu,s-\ell}\left(u,v\right) \\  \nonumber
&=\int_{\partial \text{AdS}} dP_5  \frac{\langle \mathcal{O}_{\Delta}\left(P_1\right) \mathcal{O}_{\Delta}\left(P_2\right) \mathcal{O}_{d/2+i\nu,s-\ell}\left(P_5,D_Z\right) \rangle \langle \mathcal{O}_{d/2-i\nu,s-\ell}\left(P_5,Z\right)  \mathcal{O}_{\Delta}\left(P_3\right) \mathcal{O}_{\Delta}\left(P_4\right) \rangle}{\beta_{\nu,\Delta,s-\ell}},
\end{align} 
where
\begin{equation}
b_{s-\ell}\left(\nu \right) = \frac{\nu^2 \left( g_{\phi \phi s}\right)^2 \beta_{\nu,\Delta,s-\ell}}{\pi \left(s-\ell\right)! \left(\tfrac{d}{2}-1\right)_{s-\ell}} f_{s,\ell}\left(\nu\right) \alpha_{s-\ell}\left(\nu\right) b(\Delta,\Delta,\tfrac{d}{2}+i\nu,s-\ell) b(\Delta,\Delta,\tfrac{d}{2}-i\nu,s-\ell). \label{pw}
\end{equation}
The explicit form of $ \alpha_{s-\ell}\left(\nu\right)$ is quite complicated, and is given in appendix \ref{app:tracelessexch} together with the details of its derivation. Comparing with the definition \eqref{cpwe} of the CPWE, note that the series for this single four-point exchange is truncated.

We have hence successfully computed the four-point exchange of a spin-$s$ gauge boson between two pairs of external real bulk scalars (figure \ref{fig:exchange_s}), expressing the result in the 
form of a CPWE on the boundary. This is one of the main results of the paper. In the next section, we carry out the same computation in the manifest trace gauge.

\subsection{Manifest trace gauge}
\label{Sec:FourPointExchange;Subsec:ManifestTraceGauge}
We now turn to the manifest trace gauge, in which we compute the exchange using the propagator \eqref{eq:ManifestTrace}. As with the traceless gauge in the previous section, we use the 
split representation \eqref{splitmt}, and the exchange \eqref{genexch} thus acquires the form
\begin{align} 
& \mathcal{A}_{s,\phi}\left(P_1,P_2 ; P_3,P_4\right)
\label{exchmtr}
= \left(g_{\phi \phi s}\right)^2 \int^{\infty}_{-\infty} d\nu \sum_{k=0}^{[s/2]} g_{s,k}\left(\nu\right) \frac{\nu^2}{\pi \left(s-2k\right)!\left(\tfrac{d}{2}-1\right)_{s-2k}} \int_{\partial \text{AdS}} d P_5 \\ \nonumber
& \qquad \qquad \qquad \times \int_{\text{AdS}} dX_1 \left\{ \Pi_{d/2+i\nu,s-2k}\left(X_1,P_5;\partial_{U_1},D_Z\right)\right\} (\partial_{U_1}\cdot\partial_{U_1})^k{\cal J}_s(X_1,{U_1},P_1,P_2) \\ \nonumber
& \qquad \qquad \qquad \times \int_{\text{AdS}} dX_2 \left\{ \Pi_{d/2-i\nu,s-2k}\left(X_2,P_5;\partial_{U_2},Z\right)\right\} (\partial_{U_2}\cdot\partial_{U_2})^k{\cal J}_s(X_2,{U_2},P_3,P_4). 
\end{align}
To proceed, let us first note that the $k=0$ term
\begin{equation} \label{leadingterm}
\int_{\partial \text{AdS}} d P_5 \int_{\text{AdS}} dX_1 \left\{ \Pi_{d/2+i\nu,s}\left(X_1;P_5\right)\right\}\cdot {\cal J}_s(X_1,P_1,P_2)  \int_{\text{AdS}} dX_2 \left\{ \Pi_{d/2-i\nu,s}\left(X_2;P_5\right)\right\} \cdot {\cal J}_s(X_2,P_3,P_4),
\end{equation}
(modulo pre-factors) can be readily brought into the form of the $k=0$ term in the CPWE:
 Due to tracelessness of 
$\Omega_{\nu,s}$ (manifest in the braces around the $\Pi_{d/2\pm i\nu,s}$) one can drop pure trace terms in the currents ${\cal J}_s$, represented
as $u^2(\dots)$ in \eqref{rankscur2}. Then, using integration by parts and taking into account 
that $\Omega_{\nu,s}$ is also divergence-free, one can see that the remaining terms
in the currents can be brought into the form
 $\phi(X_1,P_1) (U_1\cdot \nabla)^s \phi(X_2,P_2)$, where the derivatives act only
 on the scalar field to the right. This gives precisely the elementary vertex \eqref{simpvertex},
 which was used to define the basic three-point function \eqref{3pt}. The tensorial structure of the latter is 
 readily in the form of the $\ell=s$ term in the conformal partial wave expansion \eqref{cpwe}, and 
 the normalisation functions \eqref{bpw} of the elementary three-point 
 functions therefore contribute to the expression for $b_{s}\left(\nu\right)$.\footnote{Recall that in the definition \eqref{cpwe} of the CPWE the three-point functions have unit coefficient.}
 
 It turns out that one can do the same for the other terms in the sum \eqref{exchmtr} associated with non-zero $k$: In the formula \eqref{repeatedtr} for the traces of the currents,
 \begin{equation}
\left(\partial_u\cdot \partial_u\right)^k{\cal J}_s(x,u)
= \tau_{s,k}\left(\nu\right){\cal J}_{s-2k}(x,u) + (u\cdot\nabla)(\dots) +  u^2  \cdot (\dots),
\end{equation}
 as argued above, the terms $(u\cdot\nabla)(\dots) +  u^2  \cdot (\dots)$ will not contribute in the exchange. Therefore in using this formula, multiple traces of the currents can be replaced with currents of lower rank, and one can make the same argument as with \eqref{leadingterm} to place each term in \eqref{exchmtr} readily in the form as in a CPWE.

For the four-point exchange amplitude \eqref{exchmtr} we thus find,
\begin{align}
&\mathcal{A}_{s,\phi}\left(P_1,P_2 ; P_3,P_4\right) 
\label{exchmtr2}
= \left(g_{\phi \phi s}\right)^2\sum_{k=0}^{[s/2]} 4^{s-2k} \int^{\infty}_{-\infty} d\nu \; \frac{g_{s,k}\left(\nu\right)\tau_{s,k}^2\left(\nu\right)\nu^2}{\pi \left(s-2k\right)!\left(\tfrac{d}{2}-1\right)_{s-2k}} \int_{\partial \text{AdS}} dP_{5}
\\ \nonumber
& \qquad \qquad \qquad \qquad \times \int_{\text{AdS}} dX_1 \left\{ \Pi_{d/2+i\nu,s-2k}\left(X_1,\partial_{U_1}\right) \right\}  \phi(X_1,P_1)(U_1\cdot\nabla_1)^{s-2k} \phi(X_1,P_2) \\ \nonumber
& \qquad \qquad \qquad \qquad \times \int_{\text{AdS}} dX_2 \left\{ \Pi_{d/2-i\nu,s-2k}\left(X_2,\partial_{U_2}\right) \right\}  \phi(X_2,P_3)(U_2\cdot\nabla_2)^{s-2k} \phi(X_2,P_4). 
\end{align}
where each bulk integral in the sum is a three-point Witten diagram of a pair of scalars interacting with a spin-$\left(s-2k\right)$ field through the elementary vertex \eqref{simpvertex}. Therefore employing \eqref{3pt} and comparing with the form \eqref{cpwe} of the CPWE, we find that the coefficients $b_{s-2k}(\nu)$ are 
\begin{align}
\label{mtpw}
& b_{s-2k}(\nu)\\ 
& \hspace*{0.5cm}= \left(g_{\phi \phi s}\right)^2\frac{ 4^{s-2k}g_{s,k}(\nu)\tau^2_{s,k}(\nu)\beta_{\nu,\Delta,s-2k} \;\nu^2}{\pi (s-2k)! (\tfrac{d}{2}-1)_{s-2k}} b(\Delta,\Delta,\tfrac{d}{2}+i\nu,s-2k)b(\Delta,\Delta,\tfrac{d}{2}-i\nu,s-2k) \nonumber \\ \nonumber
& \hspace*{0.5cm} = \left(g_{\phi \phi s}\right)^24^{s-2k}g_{s,k}(\nu)\tau^2_{s,k}(\nu) \frac{\Gamma^2\left(\frac{3-d-2(s-2k)}{2}\right)
\Gamma^2(1-\tfrac{d}{2}-(s-2k))}{\pi^{3d/2+1}2^{2d+6(s-2k)+3}\Gamma^2(2-d-2(s-2k))
\Gamma^4(\Delta+1-\tfrac{d}{2})},
\end{align}
where $g_{s,k}$ and $\tau_{s,k}$ are given by \eqref{answertrgauge} and \eqref{beta}, respectively.

Above we have completed the computation of the exchange in the manifest trace gauge. Let us stress that in contrast to the calculation in the traceless gauge of the previous section, 
very little manipulation was required to express the final form as a CPWE (compare for example with appendix \ref{app:tracelessexch}). A good check of our results is if the computations in the two different gauges
are consistent with each other. This is verified in the next section.
\subsection{Checks}
\label{subsec:checks}
In this section we verify that the expressions obtained for the four-point exchange in both the traceless gauge (section \ref{Sec:FourPointExchange;Subsec:TracelessGauge}) and the manifest trace gauge (section \ref{Sec:FourPointExchange;Subsec:ManifestTraceGauge}) are consistent with each other. Since both calculations express the result for the four-point exchange in the form \eqref{cpwe} of a CPWE, only the coefficients \eqref{pw} and \eqref{mtpw} of the partial waves \eqref{partialwave} obtained in the respective gauges need to be compared.

In the manifest trace gauge, the only partial waves which contribute to the four-point exchange are $F_{\nu,s-2k}\left(u,v\right)$, with $k=0,1,...,\left[s/2\right]$. For consistency, the coefficients $b_{s-\ell}\left(\nu\right)$ \eqref{pw} of the partial wave $F_{\nu,s-\ell}\left(u,v\right)$ calculated in the traceless gauge must then vanish for odd $\ell$. The involved nature of the explicit result in the traceless gauge (appendix \ref{app:tracelessexch}) makes this difficult to verify for arbitrary $s$. However, using Mathematica we have checked explicitly for $s$ up to 14 that the partial wave coefficient $b_{s-\ell}\left(\nu\right)$ is indeed zero for each odd $\ell$.

We now move on to consider non-zero coefficients. For small values of $\ell$ and $k$, it is still tractable to compare results in the two gauges. In the following, we explicitly show that both gauges give the same results for $b_{2}\left(\nu\right)$ and $b_{0}\left(\nu\right)$. That is, we show that \eqref{pw} and \eqref{mtpw} agree for $\ell=0$, $k=0$, and for $\ell=2$, $k=1$:

\begin{itemize}
\item \underline{{\bf $\ell=0$ and $k=0$} }

This check is straightforward, since in both gauges this contribution to the four-point exchange arises from the contraction of 
the harmonic function $\Omega_{\nu,s}\left(x_1,x_2\right)$ with the current $\mathcal{J}_{s}\left(x\right)$ at points $x_1$ and $x_2$ --- see \eqref{exchtr} and \eqref{exchmtr}. Agreement
thus depends on whether or not the weighting functions $f_{s,0}\left(\nu\right)$ and $g_{s,0}\left(\nu\right)$ associated to $\Omega_{\nu,s}\left(x_1,x_2\right)$ in the respective bulk-to-bulk propagators coincide.
This is easily checked to be the case.

\item \underline{{\bf $\ell=2$ and $k=1$} }

For the manifest trace gauge, \eqref{mtpw} gives
\begin{align} \nonumber
& b_{s-2}\left(\nu\right)\\ \nonumber
& \hspace*{0.5cm} =\frac{  \left(g_{\phi \phi s}\right)^2 \nu^2 \beta_{\nu,\Delta,s-2}}{\pi (s-2)! (\tfrac{d}{2}-1)_{s-2}}
g_{s,1}(\nu)\tau^2_{s,1}(\nu) 4^{s-2} b(\Delta,\Delta,\tfrac{d}{2}+i\nu,s-2) b(\Delta,\Delta,\tfrac{d}{2}-i\nu,s-2)\\ \nonumber 
& \hspace*{0.5cm} = - \frac{  \left(g_{\phi \phi s}\right)^2 \nu^2 \beta_{\nu,\Delta,s-2}}{\pi (s-2)! (\tfrac{d}{2}-1)_{s-2}} b(\Delta,\Delta,\tfrac{d}{2}+i\nu,s-2) b(\Delta,\Delta,\tfrac{d}{2}-i\nu,s-2) \\ \label{mtbell2}
&\hspace*{0.5cm} \times \frac{4^{s-2} s\left(s-1\right)\left((2\Delta-d-1)(2\Delta-d+1)+\left(\nu^2+(\tfrac{d}{2}+s-1)^2\right)\right)^2}{2\left(d+2s-4\right)\left(d+2s-5\right)\left(\nu^2+\left(\tfrac{d}{2}+s-1\right)^2\right)} .
\end{align}
While in the traceless gauge,
\begin{align} \nonumber
& b_{s-2}\left(\nu\right) \\ \nonumber
&\hspace*{0.5cm} =\frac{  \left(g_{\phi \phi s}\right)^2 \nu^2 \beta_{\nu,\Delta,s-2}}{\pi (s-2)! (\tfrac{d}{2}-1)_{s-2}}\; f_{s,2} \left(\nu\right) \;\alpha_{s-2}\left(\nu\right)b(\Delta,\Delta,\tfrac{d}{2}+i\nu,s-2) b(\Delta,\Delta,\tfrac{d}{2}-i\nu,s-2) \\ \nonumber \\ \nonumber
&\hspace*{0.5cm} = \frac{  \left(g_{\phi \phi s}\right)^2 \nu^2 \beta_{\nu,\Delta,s-2}}{\pi (s-2)! (\tfrac{d}{2}-1)_{s-2}} b(\Delta,\Delta,\tfrac{d}{2}+i\nu,s-2) b(\Delta,\Delta,\tfrac{d}{2}-i\nu,s-2) \\  
& \hspace*{1cm} \times f_{s,2}\left(\nu\right) \mathcal{T}_{d/2+i\nu,s,s-2} \mathcal{T}_{d/2-i\nu,s,s-2}.
\end{align}
From the explicit expression \eqref{bigtau} for $\mathcal{T}_{d/2 \pm i\nu,s,s-2}$, it is possible to show that
\begin{align}
 & \mathcal{T}_{d/2 \pm i\nu,s,s-2} \\ 
 & \hspace*{0.5cm}= \frac{2^{s-2}\left(\left(\tfrac{d}{2}+s-2\right)^2+\nu^2\right) \left((2\Delta-d-1)(2\Delta-d+1)+\left(\nu^2+(\tfrac{d}{2}+s-1)^2\right)\right)}{\left(d+2s-3\right)}. \nonumber
\end{align} 
Together with 
\begin{equation}
f_{s,2}\left(\nu\right) = -\frac{s\left(s-1\right)\left(d+s-3\right)^2}{2\left(d+2s-4\right)\left(d+2s-5\right)} \frac{1}{\left(\nu^2+\left(\tfrac{d}{2}+s-2\right)^2\right)^2} \frac{1}{\nu^2+\left(\tfrac{d}{2}+s-1\right)^2},
\end{equation}
one finds agreement with the result \eqref{mtbell2} in the manifest trace gauge.
\end{itemize}

\subsection{On improvements}
\label{subsec:improvements}
 \begin{figure}[h]
 \centering
\includegraphics[width=1\linewidth]{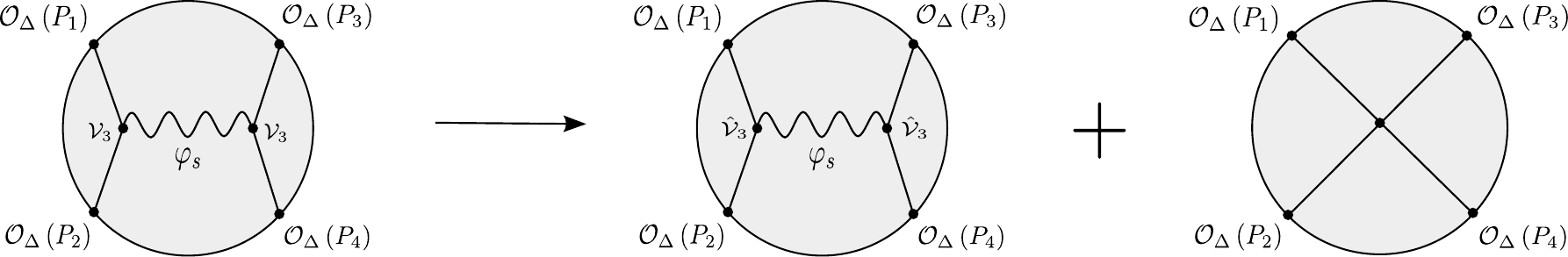} 
 \caption{The presence of trivial terms in the cubic vertex $\mathcal{V}_{3}$ causes the decomposition of the exchange amplitude into a exchange governed by the on-shell vertex $\mathcal{\hat V}_{3}$, and a quartic scalar contact interaction.}
\label{fig:improvement}
\end{figure}
It is well-known that cubic vertices involving spin-$s$ field and two scalar fields
 are defined uniquely up to terms that vanish
on the free shell. These terms in turn can be removed by field redefinitions, which is why they
are usually neglected in the analysis of cubic interactions. However, in the context of AdS/CFT, free 
bulk fields can be uniquely fixed at the level of two-point functions. Hence,
such field redefinitions would no be longer allowed. This means that such trivial on-free-shell
vertices should also be taken into account. In this section we briefly discuss the
effect of these cubic vertices on four-point exchanges.

Given a rank-$\left(s-2\right)$ conserved current ${\cal J}_{s-2}$, one can show that the quantity
\begin{equation}
\label{imp1}
{\cal D}{\cal J}_{s-2}\equiv (u^2\Box -(u\cdot \nabla)^2-u^2(u\cdot\partial_u+1)(u\cdot\partial_u +d)){\cal J}_{s-2}
\end{equation}
is also conserved.\footnote{Moreover, the cubic coupling to the massless spin-$s$ field that ${\cal D}{\cal J}_{s-2}$
 generates is trivial on-shell, which follows from the fact that the on-shell cubic vertex
$s-0-0$ is unique.}
This means that ${\cal D}{\cal J}_{s-2}$ may be viewed as an improvement term
to the rank-$s$ conserved current ${\cal J}_s$. More generally, one can consider an improvement of
the form
\begin{equation}
\label{imp2}
{\cal J}_s \quad \rightarrow \quad {\cal J}'_s= {\cal J}_s + \sum_{n=1}^{[s/2]} \gamma_n{\cal D}^n {\cal J}_{s-2n},
\end{equation}
where $\gamma_n$ are arbitrary coefficients.

To illustrate the effect of such improvement terms on the four-point exchange, let us 
consider the case of the graviton exchange in the manifest trace gauge. Here, the propagator reads
\begin{equation}
\label{gravprop}
\Pi_2=\int^{\infty}_{-\infty} \frac{d\nu}{\nu^2+\frac{\:d^2}{4}}\Omega_{\nu,2}+
u_1^2 u_2^2\int^{\infty}_{-\infty} \frac{d\nu}{\nu^2+(\tfrac{d}{2}+1)^2} \Omega_{\nu,0},
\end{equation}
which couples to a rank-2 conserved current, for which \eqref{repeatedtr} gives
\begin{equation}
(\partial_u\cdot \partial_u){\cal J}_2= \tau_{2,1}{\cal J}_0,
\end{equation}
where
\begin{equation}
\label{beta21}
\tau_{2,1}= (2\Delta-d-1)(2\Delta-d+1)+\left(\nu^2+(\tfrac{d}{2}+1)^2\right).
\end{equation}
Then the graviton exchange is then
\begin{align} \label{imps2}
\mathcal{A}_{2,\phi}\left(P_1,P_2; P_3,P_4\right) 
=\int^{\infty}_{-\infty} \frac{d\nu}{\nu^2+\frac{\:d^2}{4}}\Omega_{\nu,2} {\cal J}_2{\cal J}_2+
\int^{\infty}_{-\infty} \frac{d\nu}{\nu^2+(\tfrac{d}{2}+1)^2} \tau^2_{2,1}(\nu)\Omega_{\nu,0} {\cal J}_0{\cal J}_0.
\end{align}
Now consider the effect of an improvement \eqref{imp2} to the spin-2 current ${\cal J}_2$,
\begin{equation}
{\cal J}'_2= {\cal J}_2 +\gamma_1 {\cal D}{\cal J}_0.
\end{equation}
Since the improvement term is precisely of the form \eqref{equivalence1}, it does not
contribute to the traceless and transverse part of the exchange (the first term in equation \eqref{imps2}).\footnote{It should be noted that this does not mean that the effect of improvements is not observed in the traceless \emph{gauge}, for there the bulk-to-bulk propagators are not transverse. In this case, one would expect the second term in \eqref{imp1} to contribute.} On the other hand, it does contribute
to the trace part of the exchange:
\begin{equation} \label{ibp}
(\partial_u\cdot \partial_u){\cal D}{\cal J}_0=d(\Box-d-1){\cal J}_0 =
-d\left(\nu^2+(\tfrac{d}{2}+1)^2\right){\cal J}_0,
\end{equation}
where in the second equality we implicitly integrated by parts, applying the equation of motion \eqref{harmoniceq} of the harmonic function 
from the bulk-to-bulk propagator. The expression in brackets is the same as the denominator of the trace part
of the propagator  in \eqref{gravprop}. This is a simple consequence of the fact that
the vertices generated by improvements of currents are trivial on-free-shell.
By choosing $\gamma_1$ properly one can eliminate the $\nu^2$-dependence in 
$\tau_{2,1}$ \eqref{beta21}, so that its improved value  is
\begin{equation}
\label{beta21pr}
\tau'_{2,1}= (2\Delta-d-1)(2\Delta-d+1).
\end{equation}
This vanishes when the scalar field is conformal, which is a manifestation of the fact
that for a conformal scalar the currents can be made traceless.

More generally, a spin-$s$ current has $[s/2]$ independent improvements \eqref{imp2}.
It is not hard to see that for the $k$-fold trace of the improved current ${\cal J}'_s$,
only the improvements with $n$ $\le$ $k$  contribute non-trivially while the remaining
improvements produce terms of the form \eqref{equivalence1}. In the same time,
the $k$-fold trace of the original current ${\cal J}_s$
is given by \eqref{repeatedtr} and \eqref{beta}. All terms with $m\ge 1$
are polynomials in $\nu^2$ of the degree up to $k$, and moreover they vanish on the free shell.
This follows from matching the zeros of terms with $m\ge 1$ in  \eqref{beta} 
with the poles of the propagator \eqref{answertrgauge}.
It is therefore tempting to conjecture that the $k$ improvements that contribute
to $k$-fold trace can be chosen, analogously to the example of the graviton exchange
considered above, to remove the $\nu^2$-dependent terms in $\tau_{s,k}$. This would lead to an improved value in \eqref{repeatedtr}
\begin{equation}
\tau'_{s,k}\left(\nu\right)=2^{2k}
(\Delta-\tfrac{d}{2}-k+m+1/2)_{k}.
\end{equation}
As expected, this vanishes for conformal scalar.

For any spin-$s$ we have just shown that by properly improving the current we can
 completely remove trivial on-free-shell contributions to the $k$-fold trace part of the exchange
 for any given $k$. However, as one can clearly see from lower-spin examples,
 the way that one should improve the same spin-$s$ current
  to achieve this goal differs for different $k$.
 Therefore, such on-free-shell trivial contributions cannot be removed for 
 every $k$ in the exchange \eqref{exchmtr} at the same time --- unless the scalar field is conformal. 

Now that we have established that the exchange computation is not blind to improvement terms, let us now consider the implication of their presence. 
In the exchange computation, the cubic vertex in which the spin-$s$ bulk-to-bulk propagator joins with the scalar bulk-to-boundary 
propagators is off-shell. This means that in equality \eqref{ibp}, one also picks up a Dirac delta function coming from the propagator equation
\begin{equation}
 \Box \Pi_{s} + ... = \delta^{d+1}\left(x_1,x_2\right),
 \end{equation}
  when one integrates by parts. The exchange therefore decomposes into an exchange 
 governed by the on-shell part of the vertices, and a contact diagram (see figure \ref{fig:improvement}). In particular, for the currents bilinear in the scalar these are quartic scalar contact diagrams. This suggests that there is some inter-play between 
trivial cubic vertices and quartic contact interactions. This is therefore an issue that needs to be considered when trying to extract the form of 
the quartic scalar contact interaction holographically. We discuss this point further in the conclusion.

 In this section we studied the effect on the four-point exchange produced by cubic vertices that vanish
 when on-shell with respect to the free higher-spin equations of motion. Similarly, there exist vertices that vanish when the
 scalar satisfies the free equations of motion. Their contributions have not been taken into account in our
 analysis, which can be seen from equations \eqref{repeatedtr} and \eqref{traceofXEcurrent} which hold only on-scalar-free-shell. 
 We leave studying the effect of such vertices on the four-point exchange for future research. 
 
\section{Conclusion and outlook}
\label{sec:concl}
The results established in this work constitute the first step towards our larger goal of probing locality properties of bulk interactions
 through holography. Within this programme, a simple case that one can first consider is to resolve the nature of the quartic contact interaction of 
 the real scalar in the minimal sector of higher-spin theory on AdS$_{d+1}$. The idea of the approach is quite simple: To compute the four-point function of the scalar singlet bilinear operator of the $d$-dimensional free $O\left(N\right)$ vector model holographically in the metric-like formulation, it is required to know this vertex for the contact diagram, together with results for the four-point exchange diagrams (figure \ref{fig:total4pt}). On the other hand, in its absence one should be able to deduce its form if in possession of the CFT result and the total contribution from the exchange computations. 
 
To manoeuvre ourselves towards this position, in the present paper we computed the four-point exchange of a single spin-$s$ gauge boson between two pairs of real bulk scalars (figure \ref{fig:exchange_s}) of arbitrary mass, in arbitrary dimension and in two different gauges. In order to do so, we first needed to establish the complete massless spin-$s$ bulk-to-bulk propagators in the metric-like form, to supplement the known results for bulk-to-boundary propagators and cubic vertices. These were derived in three different gauges: de Donder gauge, a traceless gauge and a manifest trace gauge. The latter can be derived from either of the former two by shifting to an unconstrained gauge parameter, which allowed the dropping of gradient terms. In addition to providing a useful gauge in which to compute the exchange, it also served as a good consistency check for the propagators. 

The spin-$s$ bulk-to-bulk propagators were derived in a basis of AdS harmonic functions, whose split representation in terms of integral products of bulk-to-boundary propagators (section \ref{subsec:split}) allowed the resultant exchange amplitudes to be written as a conformal partial wave decomposition on the boundary, ready to compare with the analogous form of the CFT result. We carried out this computation of the exchange amplitude in both the traceless gauge and the manifest trace gauge, verifying for explicit examples that the two gauges yield the same results. We note the particular simplicity of the manifest trace gauge, which required comparably little manipulation to be brought into the form of a conformal partial wave expansion. This was facilitated by combining its manifest trace structure with an identity we derived for multiple traces of the currents, which expresses them in terms of currents of lower rank \eqref{repeatedtr}. 

However, before these results can be applied to extract the quartic scalar vertex, some issues remain that we did not address in the present paper. First, one must take into account the exchange contributions from the remaining exchange channels, and also for each spin $s=0, 2, 4, ...$ appearing in the minimal sector of the theory.\footnote{Moreover, the series should be summed in a manageable form. The corresponding summations of current exchanges can be performed in flat spacetime and the final result takes a very compact form \cite{Bekaert:2009ud}, suggesting that the (A)dS case should be tractable as well.} 
 Second, as discussed in the introduction, at the cubic level there is a degree of arbitrariness in the interactions, which manifest themselves in current interactions as ``improvements'' to genuine Noether currents. Such vertices are often neglected in the literature (with the important exception of \cite{Boulanger:2008tg} where the role of the ``Born-Infeld tail'' in Vasiliev's theory was also discussed), however in section \ref{subsec:improvements} we studied improvements to the currents entering the cubic vertices of the exchange, and demonstrated that despite such terms vanishing on the free-shell, they do play a role in exchange computations. In particular, we observed that these trivial contributions to the cubic vertices generate contact terms, such that the exchange amplitude decomposes into an exchange governed by the on-shell cubic vertices and a contact diagram (figure \ref{fig:improvement}). These contributions therefore need to be carefully considered when trying to reconstruct the quartic scalar vertex holographically: It could be the case that the ambiguity in the cubic vertices can be fixed at the level of three-point functions, for example by comparing to contact terms in the corresponding CFT result. Whether this is possible needs to be clarified, but it would imply that the form of the quartic contact vertex is fixed, and could therefore be inferred holographically at the level of four-point functions. Should this turn out not to be the case, it opens up the possibility that parts of the quartic contact vertex could be absorbed off-shell in the current exchange by trivial vertices, and vice-versa. More speculatively, perhaps a would-be non-local quartic vertex could be tamed by absorbing parts of the vertex in trivial vertices of the current exchange, conceivably to the extent that the quartic vertex becomes local. We will address such questions in future research.

\acknowledgments

We thank M. Costa, S. Giombi, E. Joung, E. Skvortsov, D. Sorokin, M. Taronna and A. Tseytlin for discussions. In particular, X.B. and D.P. are grateful to N. Boulanger and E. Meunier for useful exchanges on propagators at an early stage of this project. X.B. also acknowledges D.~Francia for several discussions on locality and the relevance of on-shell trivial cubic vertices at quartic level. J.E. and C.S. are grateful to H. Osborn for useful correspondence on conformal partial wave expansions. C.S. would further  like to thank M. Flory, M. Fuchs, D. Herschmann and S. Steinfurt for continuous discussions, and V. Gon\c{c}alves for a useful exchange on the split representation of the massive spin-2 bulk-to-bulk propagator.

The research of X.B. was supported by the Russian Science Foundation grant 14-42-00047 in association with Lebedev Physical Institute. The work of D.P. was supported by the  DFG grant HO 4261/2-1 ``Generalized dualities relating
gravitational theories in 4-dimensional Anti-de Sitter space with 3-dimensional conformal field
theories''. The work of J.E. and C.S. was partially supported by the European Science Foundation Holograv network (Holographic methods for strongly coupled systems). J.E. and C.S. also acknowledge the kind hospitality of le Laboratoire de Math\'ematiques et Physique Th\'eorique (LMPT), where part of this work was done. Reciprocally, X.B. thanks for warm hospitality the Max-Planck-Institut f\"ur Physik of Munich (Werner-Heisenberg-Institut), where this collaboration was initiated.

\appendix

\section{Operations with ambient tensors}
\label{appendix:operators}
To arrive at the results in this paper, we mainly rely on an operator formalism where index contractions and symmetrisation of indices (including tracelessness) are realised in terms of auxiliary vectors. Tensor operations are then translated into an operator calculus, which simplifies manipulations. We present the essentials for the ambient formalism in this appendix.

\subsubsection*{Contractions}
Here, the contraction of two rank-$r$ symmetric tensors $T\left(X,U\right)$ and $S\left(X,U\right)$ is simply
\begin{equation}
T_{A_1...A_r}\left(X\right) S^{A_1...A_r}\left(X\right) = \frac{1}{r!} T\left(X,\partial_{U} \right) S\left(X,U\right) \Big|_{U=0} = \frac{1}{r!} S\left(X,\partial_{U} \right) T\left(X,U\right)\Big|_{U=0}.
\end{equation}
Note that throughout this paper we drop the explicit setting of the auxiliary vector to zero when expressing tensor contractions through generating functions. This applies for all auxiliary vectors $u$, $w$, $U$, $W$ and $Z$.

Analogous to the above, the symmetric and \emph{traceless} contraction of $T\left(X,U\right)$ and $S\left(X,U\right)$ is 
\begin{equation}
T_{\left\{A_1...A_r\right\}}\left(X\right) S^{\left\{A_1...A_r\right\}}\left(X\right) = \frac{1}{r!\left(\tfrac{d}{2}-\frac{1}{2}\right)_r} T\left(X, K \right) S\left(X,W\right) = \frac{1}{r!\left(\tfrac{d}{2}-\frac{1}{2}\right)_r} S\left(X, K \right) T\left(X,W\right),
\end{equation}
by virtue of \eqref{sandtcads}.\\ \\
$K_{A}$ is given explicitly by
\begin{align} \label{K}
 K_{A} & = \frac{d-1}{2} \left( \frac{\partial}{\partial W^{A}} + X_{A} \left(X \cdot \frac{\partial}{\partial W} \right)\right) + \left(W \cdot \frac{\partial}{\partial W} \right)\frac{\partial}{\partial W^{A}} \\ \nonumber \\ \nonumber
 &\hspace{0.7cm}+ X_{A}  \left(W \cdot \frac{\partial}{\partial W}\right)  \left(X \cdot \frac{\partial}{\partial W}\right) - \frac{1}{2} W_{A} \left(\frac{\partial^{2}}{\partial W \cdot \partial W} +  \left(X \cdot \frac{\partial}{\partial W}\right) \left(X \cdot \frac{\partial}{\partial W}\right)\right). \nonumber
\end{align}
If the contracted tensors are already traceless and transverse to the AdS manifold, then $K_{A}$ reduces to 
\begin{equation}
K_{A}=\left(\tfrac{d}{2}-\tfrac{1}{2}+W \cdot \frac{\partial}{\partial W}\right) \frac{\partial}{\partial W^{A}}.
\end{equation}
For the symmetric and traceless contraction of $r$ vectors $T_A$ and $S_A$, the result can be expressed in terms of a Gegenbauer polynomial $C^{d/2-\frac{1}{2}}_{r}\left(t\right)$:
\begin{equation}
\frac{\left(K \cdot S\right)^{r}\left(W \cdot T\right)^{r}}{r! \left(\tfrac{d}{2}-\frac{1}{2}\right)_{r}} = B^{\left\{A_1\right.}...B^{\left.A_r\right\}} D_{\left\{A_1\right.}...D_{\left.A_r\right\}} = \frac{r!\left(B^2 D^2\right)^{\frac{r}{2}}}{2^r \left(\tfrac{d}{2}-\frac{1}{2}\right)_{r}} C^{d/2-\frac{1}{2}}_{r}\left(t\right),
\end{equation}
where
\begin{equation}
B_{A} = S_{A}+\left(X \cdot S\right) X_{A}, \quad D_{A} = T_{A}+\left(X \cdot T\right) X_{A}, \quad t = \frac{B \cdot D + \left(B \cdot X\right)\left(D \cdot X\right)}{\sqrt{\left(\left(B \cdot X\right)^2+ B^2\right)\left(\left(D \cdot X\right)^2+ D^2\right)}}.
\end{equation}
This can be extended to $r$ contractions of more than two vectors, for example the symmetric and traceless contraction of $r$ vectors $T_{A}$, with $n$ vectors $S^A_1$ and $r-n$ vectors $S^A_2$ can be computed via
\begin{equation} \label{multcont}
\frac{\left(K \cdot S_1\right)^{n}\left(K \cdot S_2\right)^{r-n}\left(W \cdot T\right)^{r}}{r! \left(\tfrac{d}{2}-\frac{1}{2}\right)_{r}} =\frac{1}{\left(r-n+1\right)_{n}} \left(S_1 \cdot \frac{\partial}{\partial S_2} \right)^{n} \frac{\left(K \cdot S_2\right)^{r}\left(W \cdot T\right)^{r}}{r! \left(\tfrac{d}{2}-\frac{1}{2}\right)_{r}}.
\end{equation}
We make use of \eqref{multcont} in the computation of the four-point exchange in the traceless gauge, with details given in appendix \ref{app:tracelessexch}.

\subsubsection*{Differential operators}
Recall from \eqref{tldiv} that the symmetric and traceless divergence can be represented by the operator $\nabla \cdot K$, which together with the gradient $W \cdot \nabla$, Laplacian $\nabla^{2}$ 
and $\mathcal{D}_{W} = W \cdot \partial_{W}$ (which returns the spin of the tensor acted on) obeys the commutation relations
\begin{align} \label{divcim}
\left[\nabla \cdot K,W \cdot \nabla \right] &= \left(\tfrac{d}{2}-\tfrac{1}{2}+\mathcal{D}_{W}\right)\nabla^{2}-\left(\mathcal{D}_{W}^{2}+3\left(\tfrac{d}{2}-\tfrac{1}{2}\right)\mathcal{D}_{W}+
\left(\tfrac{d}{2}-\tfrac{1}{2}\right)^{2}\right)\mathcal{D}_{W}, \\ \nonumber \\ \label{nablacom}
\left[\nabla^2,W \cdot \nabla \right] &= -2\left(\tfrac{d}{2}-1+\mathcal{D}_{W}\right) W \cdot \nabla.
\end{align}
One can then further show that \cite{Costa:2014kfa}
\begin{equation} \label{i1}
\left[\nabla^2,\left(W \cdot \nabla\right)^{n} \right] = -n\left(d-1+2\mathcal{D}_{W}-n\right) \left(W \cdot \nabla\right)^{n},
\end{equation}
and 
\begin{align} \label{i2}
\left[\nabla \cdot K,\right.&\left.\left(W \cdot \nabla\right)^{n} \right] \\ \nonumber
&= \frac{n}{2} \left(W \cdot \nabla\right)^{n-1} \left(d+n+2\mathcal{D}_{W}-2\right) \left(1-n-\left(n+\mathcal{D}_{W}-1\right)\left(d+n+\mathcal{D}_{W}-2\right)+\nabla^{2}\right). \nonumber
\end{align}
These commutation relations are particularly useful in deriving the explicit form for the bulk-to-bulk propagators in de Donder and the traceless gauge, in sections \ref{Sec:Propagators;Subsec:deDonder} and \ref{Sec:Propagators;Subsec:tracelessgauge} respectively.

\section{Exchange computation in traceless gauge}
\label{app:tracelessexch}
Much of the computation in this gauge follows \cite{Costa:2014kfa}, in particular section 6 of the paper where a partial contribution to the four-point exchange of \emph{massive} bosonic higher-spin fields between scalars, coming from the traceless part of the higher-spin propagator was computed. Here we are however concerned with the complete exchange of massless higher-spin fields, which couple to two scalar fields instead in a current interaction and whose bulk-to-bulk propagators were previously unknown in the metric-like formalism. 

The expression obtained for the exchange in this gauge is very involved, as will become clear in the following. However, it should be noted that we are able to check that it is in agreement with the result in section \ref{Sec:FourPointExchange;Subsec:ManifestTraceGauge} for the manifest trace gauge. We carry out this verification in section \ref{subsec:checks}.

Repeating ourselves here for convenience, our goal is to bring the split form of the massless spin-$s$ exchange in the traceless gauge,
\begin{align} \label{atrlexch}
& \mathcal{A}_{s,\phi}\left(P_1,P_2; P_3,P_4\right) = \sum^{s}_{\ell=0} \int^{\infty}_{-\infty} d\nu f_{s,\ell}\left(\nu\right) \frac{\nu^2}{\pi \left(s-\ell \right)!\left(\tfrac{d}{2}-1\right)_{s-\ell}} \int_{\partial \text{AdS}} dP_5  \\ \nonumber
&\qquad \qquad \times \frac{g_{\phi \phi s}}{s!\left(\tfrac{d}{2}-\frac{1}{2}\right)_{s}} \int_{\text{AdS}} dX_1 \; \mathcal{J}_{s}\left(X_1,K_1;P_{1},P_{2}\right) \left(W_1 \cdot \nabla_1\right)^{\ell} \Pi_{d/2+i\nu,s-\ell}\left(X_1,P_5;W_1,Z\right) \\ \nonumber
& \qquad \qquad \times \frac{g_{\phi \phi s}}{s!\left(\tfrac{d}{2}-\frac{1}{2}\right)_{s}} \int_{\text{AdS}} dX_2 \; \mathcal{J}_{s}\left(X_2,K_2;P_{3},P_{4}\right) \left(W_2 \cdot \nabla_2\right)^{\ell} \Pi_{d/2-i\nu,s-\ell}\left(X_2,P_5;W_2,D_Z\right),
\end{align}
into the form \eqref{cpwe} of a conformal partial wave expansion.  Namely, the bulk integrals need to be expressed in terms of boundary three-point functions $\langle \mathcal{O}_{\Delta}\mathcal{O}_{\Delta}\mathcal{O}_{d/2\pm i\nu,s-\ell}\rangle$ with unit normalisation as in \eqref{canon3pt}. Unlike in the manifest trace gauge, as written above the bulk integrals are not manifestly in the desired form. One way to proceed is to simply evaluate the bulk integrals for the boundary expression, since conformal invariance guarantees that the tensorial structure is the required form -- all that remains is to extract the overall coefficient. This is the route we take in the following.

Let us focus on the single bulk integral over $X_1$ on the second line of \eqref{atrlexch}, since the same applies to $X_2$. This has the explicit form
\begin{align}
 &\hat{\mathcal{A}}_{1}\left(P_1,P_2\right) =  \frac{g_{\phi \phi s}}{s!\left(\tfrac{d}{2}-\frac{1}{2}\right)_{s}} \sum^{s}_{a=0} \frac{\left(-1\right)^a s!}{a!\left(s-a\right)!}\int_{\text{AdS}} dX_1 \\ 
 & \qquad \times  \left(K_1 \cdot \nabla_1\right)^{\ell} \Pi_{d/2+i\nu,s-\ell}\left(X_1,P_5;K_1,Z\right)
  \left(W_1 \cdot \nabla_1\right)^{a} \phi \left(X_1,P_1\right) \left(W_1 \cdot \nabla_1\right)^{s-a} \phi \left(X_1,P_2\right). \nonumber
\end{align}
The integral can be evaluated by noting that 
\begin{align}
\left(K \cdot \nabla\right)^{\ell} & \Pi_{d/2\pm i\nu,s-\ell}\left(X,P;K,Z\right) = \frac{\mathcal{C}_{d/2\pm i\nu,s-\ell} \left(d/2\pm i\nu + s-\ell\right)_{\ell}}{\left(-2 P \cdot X\right)^{d/2\pm i\nu+s}} \left(2 P \cdot K\right)^{\ell} \left( \bar X \cdot K \right)^{s-\ell}, \nonumber
\end{align}
where we introduced
\begin{equation}
\bar X_A =  Z_A\left(2 P \cdot X\right) - 2 P_A\left(Z \cdot X\right),
\end{equation}
and also
\begin{equation}
\left(W \cdot \nabla \right)^{a} \phi \left(X,P\right) = \mathcal{C}_{\Delta} \left(\Delta\right)_{a} \left(-2 P \cdot X\right)^{-\left(a+\Delta\right)} \left(2 W \cdot P\right)^{a}.
\end{equation}
Then employing the identity \eqref{multcont} for the symmetric and traceless contraction between the conserved current and $\left(W \cdot \nabla\right)^{\ell} \Pi_{d/2\pm i\nu,s-\ell}$, and expressing 
the resultant Gegenbauer polynomial in terms of the hypergeometric function $\mathstrut_2 F_{1}$, we find  
\begin{align}
 &\hat{\mathcal{A}}_{1}\left(P_1,P_2\right) = \\ \nonumber
& g_{\phi \phi s} \sum^{(1)}\frac{\left(P_{15}\right)^{m-b}}{\left(P_{25}\right)^{b-e+s-\ell}}\int_{\text{AdS}} dX_1 \frac{\left(\mathcal{C}_{\Delta}\right)^2\mathcal{C}_{d/2+i \nu,s-\ell}\left(-\bar X_1 \cdot P_1\right)^{b}\left(-\bar X_1 \cdot P_2\right)^{s-\ell-b}}{\left(-2P_1 \cdot X_1\right)^{\Delta+m}\left(-2P_2 \cdot X_1\right)^{\Delta+e}\left(-2P_5 \cdot X_1\right)^{d/2+i\nu +m +e}} \\ \nonumber  \\ \nonumber 
&+ g_{\phi \phi s} \sum^{(2)}\frac{\left(P_{15}\right)^{m-b}\left(P_{12}\right)^{g}}{\left(P_{25}\right)^{b-e+s-\ell}}\int_{\text{AdS}} dX_1 \frac{\left(\mathcal{C}_{\Delta}\right)^2\mathcal{C}_{d/2+i \nu,s-\ell}\left(-\bar X_1 \cdot P_1\right)^{b}\left(-\bar X_1 \cdot P_2\right)^{s-\ell-b}}{\left(-2P_1 \cdot X_1\right)^{\Delta+m+g}\left(-2P_2 \cdot X_1\right)^{\Delta+e+g}\left(-2P_5 \cdot X_1\right)^{d/2+i\nu +m +e}}, \\ \nonumber 
\end{align}
where we introduced
\begin{align}
\sum\limits^{(1)}=&  \left(d/2+i\nu +s-\ell\right)_{\ell} \frac{\ell ! \left(-1\right)^{\frac{s}{2}}}{2^s s! \left(\tfrac{d}{2}-\frac{1}{2}\right)_{s}} \sum^{s}_{a=0} \sum^{s-\ell}_{b=0} \sum^{s/2}_{c=0} \sum^{a}_{m=0} \sum^{2c-a}_{e=0}   \left(-1\right)^{a+c+m+e} 2^{2c+m+e} \\ \nonumber
& \times \frac{\binom{s}{a}\binom{s-\ell}{b} \binom{a}{m}\binom{2c-a}{e} m! e! \left(s-a\right)! \left(\Delta\right)_{a} \left(\Delta\right)_{s-a} \left(\tfrac{d}{2}-\frac{1}{2}\right)_{\tfrac{s}{2}+c}}{\left(m-b\right)!\left(b+e+\ell-s\right)!\left(\tfrac{s}{2}-c\right)!\left(2c-a\right)!},
\end{align}
and 
\begin{align}
& \sum^{(2)}= \left(d/2+i\nu +s-\ell\right)_{\ell} \frac{\ell ! \left(-1\right)^{\frac{s}{2}}}{2^s s! \left(\tfrac{d}{2}-\frac{1}{2}\right)_{s}} \sum^{s}_{a=0} \sum^{s-\ell}_{b=0} \sum^{s/2}_{c=0}  \sum^{a-1}_{n=0}\sum^{n}_{m=0} \sum^{2c-n}_{e=0} \sum^{a-n}_{f=1} \sum^{2k+n-a}_{g=0} \left(-1\right)^{a+c+m+e+g} 
\\ \nonumber
& \times 2^{-a+2c+m+e+g+2f+n} \frac{\binom{s}{a}\binom{s-\ell}{b} \binom{n}{m}\binom{2c-n}{e} \binom{2f+n-a}{g}\binom{a}{n} m! e! \left(s-a\right)! \left(a-n\right)! \left(\Delta\right)_{a} \left(\Delta\right)_{s-a} \left(\tfrac{d}{2}-\frac{1}{2}\right)_{\tfrac{s}{2}+c}}{\left(m-b\right)!\left(b+e+\ell-s\right)!\left(\tfrac{s}{2}-c-f\right)!\left(2c-n\right)!\left(2f+n-a\right)! \left(a-n-f\right)!}. \nonumber
\end{align}
The bulk integral can now be easily evaluated with the help of the boundary differential operator \eqref{DP},  $\mathcal{D}_{P}$. To wit,
\begin{align}
&\frac{\left(P_{15}\right)^{m-b}\left(P_{12}\right)^{g}}{\left(P_{25}\right)^{b-e+s-\ell}} \int_{\text{AdS}} dX_1 \frac{\left(\mathcal{C}_{\Delta}\right)^2\mathcal{C}_{d/2+i \nu,s-\ell}\left(-\bar X_1 \cdot P_1\right)^{b}\left(-\bar X_1 \cdot P_2\right)^{s-\ell-b}}{\left(-2P_1 \cdot X_1\right)^{\Delta+m+g}\left(-2P_2 \cdot X_1\right)^{\Delta+e+g}\left(-2P_5 \cdot X_1\right)^{d/2+i\nu +m +e}} \\ \nonumber \\ \nonumber
&= \frac{\left(P_2 \cdot \partial_U \mathcal{D}_{P_5}\right)^{s-\ell-b}\left(P_1 \cdot \partial_U \mathcal{D}_{P_5}\right)^{b}}{\left(d/2+i\nu+m+e+\ell-s\right)_{s-\ell}} \int_{\text{AdS}} dX_1 \frac{\left(P_{15}\right)^{m-b}\left(P_{25}\right)^{e-b+\ell-s} \left(P_{12}\right)^{g} \left(\mathcal{C}_{\Delta}\right)^2\mathcal{C}_{d/2+i \nu,s-\ell}}{\left(-2P_1 \cdot X_1\right)^{\Delta_1}\left(-2P_2 \cdot X_1\right)^{\Delta_2}\left(-2P_5 \cdot X_1\right)^{\Delta_5}}  \\ \nonumber \\ \nonumber
& = \left(-1\right)^{b}\frac{\left(\delta_{15}\right)_{s-\ell-b}\left(\delta_{25}\right)_{b}b\left(\Delta_1,\Delta_2,\Delta_5,0\right)}{ \left(d/2+i\nu+m+e+\ell-s\right)_{s-\ell}} \frac{\left(\mathcal{C}_{\Delta}\right)^2\mathcal{C}_{d/2+i \nu,s-\ell}}{\mathcal{C}_{\Delta_1}\mathcal{C}_{\Delta_2}\mathcal{C}_{\Delta_5}}\\ \nonumber
& \hspace{5cm} \times \frac{\left(P_{25}\left(Z \cdot P_1\right)-P_{15} \left(Z \cdot P_2\right)\right)^{s-\ell}}{\left(P_{12}\right)^{\frac{1}{2}\left(2\Delta+s-\ell-d/2-i\nu\right)}\left(P_{25}\right)^{\frac{1}{2}\left(d/2+i\nu+s-\ell\right)}\left(P_{15}\right)^{\frac{1}{2}\left(d/2+i\nu+s-\ell\right)}} \\ \nonumber \\ \nonumber
&= \left(-1\right)^{b}\frac{2^{\ell-s}\left(\frac{2\Delta-d/2+i\nu+s-\ell}{2}\right)_{m+e+g+\ell-s}\left(\frac{2\Delta-d/2-i\nu+s-\ell}{2}\right)_{g}\left(\frac{d/2+i\nu+s-\ell}{2}\right)_{b+e+\ell-s}\left(\frac{d/2+i\nu+s-\ell}{2}\right)_{m-b}}{ \left(\Delta\right)_{m+g}\left(\Delta\right)_{e+g}\left(d/2+i\nu+s-\ell\right)_{m+e+2\ell-2s}\left(d/2+i\nu+m+e+\ell-s\right)_{s-\ell}} \\ \nonumber \\ \nonumber
& \qquad \times b(\Delta,\Delta,\tfrac{d}{2}+i\nu,s-\ell) \langle \mathcal{O}_{\Delta}\left(P_1\right)\mathcal{O}_{\Delta}\left(P_2\right) \mathcal{O}_{d/2+i \nu,s-\ell}\left(P_5,Z\right) \rangle,
\end{align}
with
\begin{align}
\Delta_{1} =\Delta+m+g, & \quad \Delta_{2}= \Delta+e+g, \quad \Delta_{5}= \tfrac{d}{2}+i\nu +m+e+\ell-s \\ \nonumber
& \text{and} \qquad \delta_{ij} = \frac{1}{2}\left(\Delta_{i}+\Delta_{j} - \Delta_{k}\right).
\end{align}
Where we used \eqref{3pt} in the evaluation of the bulk integral. $\hat{\mathcal{A}}_{1}$ is thus evaluated as
\begin{align}
\hat{\mathcal{A}}_{1}\left(P_1,P_2\right) &= g_{\phi \phi s}\; \mathcal{T}_{d/2+i\nu,s,s-\ell} \;b(\Delta,\Delta,\tfrac{d}{2}+i\nu,s-\ell) \; \langle \mathcal{O}_{\Delta}\left(P_1\right)\mathcal{O}_{\Delta}\left(P_2\right) \mathcal{O}_{d/2+i \nu,s-\ell}\left(P_5,Z\right) \rangle,
\end{align}
where
\begin{align}
\mathcal{T}_{d/2+i\nu,s,s-\ell}= \mathcal{T}^{(1)}_{d/2+i\nu,s,s-\ell} + \mathcal{T}^{(2)}_{d/2+i\nu,s,s-\ell}, \label{bigtau}
\end{align}
\begin{align}
& \frac{\mathcal{T}^{(1)}_{d/2+i\nu,s,s-\ell}}{\left(d/2+i\nu+s-\ell\right)_{\ell}} = \\ \nonumber
& \frac{\ell! \left(-1\right)^{s/2}}{s! \left(\tfrac{d}{2}-\frac{1}{2}\right)_{s}} \sum^{s}_{a=0} \sum^{s-\ell}_{b=0} \sum^{s/2}_{c=0} \sum^{a}_{m=0} \sum^{2c-a}_{e=0} \left(-1\right)^{a+b+c+m+e} 
\frac{\left(\tfrac{d}{2}-\frac{1}{2}\right)_{\frac{s}{2}+c}\binom{a}{m}\binom{2c-a}{e}\binom{s-\ell}{b}\binom{s}{a} m! e! \left(s-a\right)!}{\left(\frac{s}{2}-c\right)!\left(2c-a\right)!\left(m-b\right)!\left(b+e+\ell-s\right)!} \\ \nonumber \\ \nonumber
&\times 2^{\ell-2s+2c+m+e} \frac{\left(\Delta\right)_{a}\left(\Delta\right)_{s-a}\left(\frac{2\Delta-d/2+i\nu+s-\ell}{2}\right)_{m+e+\ell-s}\left(\frac{d/2+i\nu+s-\ell}{2}\right)_{b+\ell-s}\left(\frac{d/2+i\nu+s-\ell}{2}\right)_{m-b}}{\left(\Delta\right)_{m}\left(\Delta\right)_{e} \left(d/2+i\nu+m+e+\ell-s\right)_{s-\ell}\left(d/2+i\nu+s-\ell\right)_{m+e+2\ell-2s}},
\end{align}
and
\begin{align}
&\frac{\mathcal{T}^{(2)}_{d/2+i\nu,s,s-\ell}}{\left(d/2+i\nu+s-\ell\right)_{\ell}} = \frac{\ell! \left(-1\right)^{s/2}}{s! \left(\tfrac{d}{2}-\frac{1}{2}\right)_{s}} \sum^{s}_{a=0} \sum^{s-\ell}_{b=0} \sum^{s/2}_{c=0} \sum^{a-1}_{n=0} \sum^{n}_{m=0} \sum^{2c-n}_{e=0} \sum^{a-n}_{f=1}\sum^{2f+n-a}_{g=0} \left(-1\right)^{a+b+c+m+e+g} \\ \nonumber
& \times \frac{2^{\ell-2s-a+2c+m+e+g+2f+n} \left(\tfrac{d}{2}-\frac{1}{2}\right)_{\frac{s}{2}+c}\binom{n}{m}\binom{2c-n}{e}\binom{s-\ell}{b}\binom{s}{a} \binom{2f+n-a}{g}\binom{a}{n}m! e! \left(s-a\right)!\left(a-n\right)!}{\left(s/2-c-f\right)!\left(2c-n\right)!\left(m-b\right)!\left(b+e+\ell-s\right)!\left(2f+n-a\right)!\left(a-n-f\right)!} \\ \nonumber \\ \nonumber
&\times  \frac{\left(\Delta\right)_{a}\left(\Delta\right)_{s-a} \left(\frac{2\Delta-d/2+i\nu+s-\ell}{2}\right)_{m+e+g+\ell-s}\left(\frac{2\Delta-d/2-i\nu+s-\ell}{2}\right)_{g}\left(\frac{d/2+i\nu+s-\ell}{2}\right)_{b+\ell-s}\left(\frac{d/2+i\nu+s-\ell}{2}\right)_{m-b}}{\left(\Delta\right)_{m+g}\left(\Delta\right)_{e+g} \left(d/2+i\nu+m+e+\ell-s\right)_{s-\ell}\left(d/2+i\nu+s-\ell\right)_{m+e+2\ell-2s}}.
\end{align}
We have thus achieved the following decomposition for the spin-$s$ exchange in the traceless gauge
\begin{align} \label{appbexch}
&\mathcal{A}_{s,\; \phi}\left(P_1,P_2; P_3,P_4\right) = \sum^{s}_{\ell=0} \int^{\infty}_{-\infty} d\nu \; b_{s-\ell}\left(\nu\right) \\ \nonumber
& \times  \int_{\partial \text{AdS}} dP_5 \;\frac{\langle \mathcal{O}_{\Delta}\left(P_1\right)\mathcal{O}_{\Delta}\left(P_2\right) \mathcal{O}_{d/2+i \nu,s-\ell}\left(P_5,D_Z\right) \rangle \langle \mathcal{O}_{d/2-i\nu,s-\ell}\left(P_5,Z\right)\mathcal{O}_{\Delta}\left(P_3\right) \mathcal{O}_{\Delta}\left(P_4\right) \rangle}{\beta_{\nu,\Delta,s-\ell}},
\end{align}
where
\begin{align} \nonumber
& b_{s-\ell}\left(\nu\right) = \frac{\nu^2 \left( g_{\phi \phi s}\right)^2 \beta_{\nu,\Delta,s-\ell}}{\pi \left(s-\ell\right)! \left(\tfrac{d}{2}-1\right)_{s-\ell}} f_{s,\ell}\left(\nu\right) \alpha_{s-\ell}\left(\nu\right) b(\Delta,\Delta,\tfrac{d}{2}+i\nu,s-\ell) b(\Delta,\Delta,\tfrac{d}{2}-i\nu,s-\ell) \\ 
&  =\frac{\nu^2 \left( g_{\phi \phi s}\right)^2 \beta_{\nu,\Delta,s-\ell} }{\pi \left(s-\ell\right)!  \left(\tfrac{d}{2}-1\right)_{s-\ell}}f_{s,\ell}\left(\nu\right)\;\mathcal{T}_{d/2+i\nu,s,s-\ell} \; \mathcal{T}_{d/2-i\nu,s,s-\ell}\; b(\Delta,\Delta,\tfrac{d}{2}+i\nu,s-\ell)\; b(\Delta,\Delta,\tfrac{d}{2}-i\nu,s-\ell),
\end{align}
and the explicit form of $\beta_{\nu,\Delta_i,\ell}$ is given by
\begin{align} \label{beta_costa}
\beta_{\nu,\Delta_i,\ell} &= \frac{2^{3-\ell}\pi^{d/2+1}\Gamma\left(i\nu\right)\Gamma\left(-i\nu\right)\left(\tfrac{d}{2}-i\nu-1\right)_{\ell}\left(\tfrac{d}{2}+i\nu-1\right)_{\ell}}{\Gamma\left(\frac{\Delta_1+\Delta_2-d/2-i\nu+\ell}{2}\right)\Gamma\left(\frac{d/2+i\nu+\ell+\Delta_1-\Delta_2}{2}\right)\Gamma\left(\frac{d/2+i\nu+\ell+\Delta_2-\Delta_1}{2}\right)\Gamma\left(\frac{\Delta_1+\Delta_2-d/2+i\nu+\ell}{2}\right)}  \\ \nonumber
& \times \frac{1}{\Gamma\left(\frac{\Delta_3+\Delta_4-d/2-i\nu+\ell}{2}\right)\Gamma\left(\frac{d/2-i\nu+\ell+\Delta_3-\Delta_4}{2}\right)\Gamma\left(\frac{d/2-i\nu+\ell+\Delta_4-\Delta_3}{2}\right)\Gamma\left(\frac{\Delta_3+\Delta_4-d/2+i\nu+\ell}{2}\right)}.
\end{align}
Let us stress that despite the complexity of the expression \eqref{appbexch} for the exchange in this gauge, in section \ref{subsec:checks} we check for explicit examples that it agrees with result \eqref{mtpw} obtained in the manifest trace gauge. 

\section{Single trace of the currents}
\label{apptraceofcur}
In this appendix, we show how a single trace of the unconstrained current \eqref{ambcur2} of a given rank can be expressed in terms of unconstrained currents of lower ranks. The formula that we will find generalises 
\begin{equation}
\label{BBVDcons}
(\partial_u\cdot \partial_u)\;I(x,u)=\left(-\Box+ 4M^2\right){ I}(x,u),
\end{equation}
from the flat space to AdS.

It is easy to obtain
\begin{equation*}
(\partial_X\cdot\partial_X+\partial_U\cdot\partial_U){\cal J}(X,U)= 0,
\end{equation*}
where it was taken into account that the ambient scalar $\Phi$ is massless \eqref{masslessamb}.
Then the (A)dS trace gives
\begin{align}
\notag
(\partial_u\cdot\partial_u){\cal J}(X,U)&\equiv\left(\partial_U\cdot\partial_U+(X\cdot \partial_U)^2\right){\cal J}(X,U)\\
\label{trcur1} \\ \nonumber
&=\;
\left(-\partial_X\cdot\partial_X+(X\cdot \partial_U)^2\right){\cal J}(X,U).
\end{align}
The next step is to rewrite the RHS of \eqref{trcur1} in terms of intrinsic (A)dS covariant 
operators: Laplacian $\Box$, rank of the current $(U\cdot\partial_U)$, multiplication by the 
metric $U^2$ and the covariant gradient
$(U\cdot\nabla)$. 
The first term in \eqref{trcur1} can be related to the Laplacian, which in ambient
notations reads
\begin{align}
\label{laplacian}
\Box=\partial_X\cdot\partial_X+(X\cdot \partial_X)&\left((X\cdot \partial_X)+d-1\right)
-\;(U\cdot& \partial_U) +2(U\cdot\partial_X)(X\cdot\partial_U)+U^2(X\cdot\partial_U)^2.
\end{align}
Then we eliminate $(X\cdot \partial_U)$ in favour of $(U\cdot \partial_X)$, using
that
\begin{equation*}
(X\cdot \partial_U)+(U\cdot \partial_X)=X^+\partial_+-X^-\partial_-.
\end{equation*}
Hence, 
\begin{equation}
\label{trcur2}
(X\cdot \partial_U){\cal J}= (\Delta_+-\Delta_--(U\cdot \partial_X)){\cal J}.
\end{equation}
Let us note that the identity above can be used only when $(X\cdot \partial_U)$
acts directly on the current. So, for example,
\begin{align*}
(X\cdot \partial_U)^2{\cal J}&=(X\cdot \partial_U) (\Delta_+-\Delta_--(U\cdot \partial_X)){\cal J}\\
&=(\Delta_+-\Delta_--(U\cdot \partial_X))(X\cdot \partial_U) {\cal J}-[(X\cdot \partial_U),(U\cdot \partial_X)]
{\cal J}\\
&=(\Delta_+-\Delta_--(U\cdot \partial_X))^2{\cal J}-(X\cdot \partial_X-U\cdot \partial_U){\cal J}.
\end{align*}
$(X\cdot\partial_X)$ can be eliminated through $(U\cdot \partial_U)$ \eqref{homcur},
\begin{equation}
(X\cdot\partial_X){\cal J}=-((U\cdot\partial_U)+d){\cal J}.
\end{equation}

Finally, we express $(U\cdot \partial_X)$ in terms of covariant gradients $(U\cdot \nabla)$, 
using
\begin{equation*}
(U\cdot \nabla)= U^2 (X\cdot\partial_U)+(U\cdot \partial_X).
\end{equation*}
Together with \eqref{trcur2}, this yields
\begin{equation}
(U\cdot \partial_X){\cal J}=\frac{1}{1-U^2}\left((U\cdot\nabla)-U^2(\Delta_+-\Delta_-)\right){\cal J},
\end{equation}
where the fraction should be understood as a power series
\begin{equation*}
\frac{1}{1-U^2}=1+U^2+U^4+\dots.
\end{equation*}
Let us stress again that the formula above can be used only when $(U\cdot \partial_X)$
acts directly on the current. In particular, 
\begin{align*}
(U\cdot \partial_X)^2{\cal J}=\left(\frac{1}{1-U^2}\left((U\cdot\nabla)-U^2(\Delta_+-\Delta_-)\right)\right)^2{\cal J}
-\frac{U^2}{1-U^2}(X\partial_X-U\partial_U){\cal J}.
\end{align*}

Therefore, in the end one finds for the trace
\begin{align}
\label{traceofXEcurrent}
\notag
\left(\partial_u\cdot\partial_u\right){\cal J}
=
\Big(-\Box+(u\cdot\partial_u+1)(&u\cdot\partial_u+d)\\
&+\;\frac{1}{1-u^2}\left((\Delta_+-\Delta_-)^2-(u\cdot\nabla)^2\right) \Big){\cal J}.
\end{align}

\section{Multiple traces of the currents}
\label{apptraceofcur1}

In this appendix we compute $\tau_{s,k}\left(\nu\right)$ in \eqref{beta}. 
To this end, we take repeated traces of \eqref{traceofXEcurrent}.
This allows to express, say, a $k-$fold trace of rank-$s$ current ${\cal J}_{s}$ in terms of lower degree traces of lower-spin currents 
\begin{eqnarray}
\label{schematically}
(\partial_u\cdot\partial_u)^{k}{\cal J}_{s}\quad \rightarrow \quad
(\partial_u\cdot\partial_u)^{k-1}{\cal J}_{s-2}, \quad(\partial_u\cdot\partial_u)^{k-2}{\cal J}_{s-4},
\quad\dots,\quad
{\cal J}_{s-2k}
\end{eqnarray}
and terms of the form
\begin{equation}
\label{tracesandgrad}
(u\cdot\nabla)(\dots) \quad \text{or} \quad  u^2  \cdot (\dots).
\end{equation}
Then  \eqref{schematically} can be applied to each term on the right hand side 
of \eqref{schematically}. Doing this repeatedly one can eliminate all the traces
from the right hand side, thereby expressing $(\partial_u\cdot\partial_u)^{k}{\cal J}_{s}$
in terms of ${\cal J}_{s-2k}$ and terms of the form \eqref{tracesandgrad}.
Our main goal is to compute the exchange, so it is
enough to know $(\partial_u\cdot\partial_u)^{k}{\cal J}_{s}$ modulo terms
which vanish upon contraction against traceless and divergencefree $\Omega_{s-2k}$.
These are precisely the terms in \eqref{tracesandgrad}. Further on we will often
drop such terms where they are unimportant. Equalities that hold modulo
these terms will be denoted by ``$\sim$''.

We will need the following useful identities
\begin{align}
\notag
(\partial_u\cdot \partial_u)^i\left(u^2\right)^k {\cal J}_l
=\sum_{n=0}^{i}&4^{i-n}\frac{i!}{n!(i-n)!}(k-i+n+1)_{i-n}\\
&\times\;((d+1)/2+k+l-i)_{i-n}
\label{appb0}
\left(u^2\right)^{k-i+n}\left(\partial_u\cdot\partial_u\right)^n{\cal J}_l,
\end{align}
and
\begin{align}
\notag
(\partial_u\cdot\partial_u)^m(u\cdot\nabla)^2{\cal J}_k=
2m\cdot\frac{k-2m+2}{k-2m+3}&\;(\partial_u\cdot \nabla)( u\cdot\nabla)(\partial_u\cdot\partial_u)^{m-1}{\cal J}_k
\\
\label{traceofgrad}
+\;
2m\;\Box\; (\partial_u&\cdot\partial_u)^{m-1}{\cal J}_k
+
(u\cdot \nabla)^2(\partial_u\cdot\partial_u)^m{\cal J}_k.
\end{align}
Here and throughout we use that the current is conserved.
To commute divergence and gradient in \eqref{traceofgrad}, we employ
\begin{equation}
\label{comdivandgrad}
\frac{1}{n+1}(\partial_u\cdot \nabla)(u\cdot\nabla){\cal J}_n=
\frac{1}{n}(u\cdot\nabla)(\partial_u\cdot \nabla){\cal J}_n-(n+d){\cal J}_n+\frac{1}{n}
u^2(\partial_u\cdot\partial_u){\cal J}_n,
\end{equation}
which entails
\begin{equation}
\label{appb1}
(\partial_u\cdot\partial_u)^m(u\cdot\nabla)^2{\cal J}_k\sim2m\left(\Box-(k-2m-2)(k-2m+d+1) \right)(\partial_u\cdot\partial_u)^{m-1}{\cal J}_k.
\end{equation}
With these formulas at hand we are prepared to compute the $i$-fold trace
$(\partial_u\cdot\partial_u)^{i}$ of the both sides of \eqref{traceofXEcurrent}.
Using \eqref{appb0}, keeping only terms with $k-i+n=0$, we obtain 
\begin{align*}
(\partial_u\cdot\partial_u)^{i+1}{\cal J}_{s+2}\sim-\;\Box\; (&\partial_u\cdot\partial_u)^i{\cal J}_s+
(s+1)(s+d)(\partial_u\cdot\partial_u)^i{\cal J}_s\\
+
\sum_{k=0}^i\frac{i!}{(i-k)!}&4^k \left((d+1)/2+s-i-k\right)_k
\\
&\times\;
(\partial_u\cdot\partial_u)^{i-k}\big((\Delta_+-\Delta_-)^2{\cal J}_{s-2k}-(u\cdot\nabla)^2{\cal J}_{s-2k-2}\big).
\end{align*}
Employing \eqref{appb1} we find
\begin{align}
\notag
(\partial_u\cdot\partial_u)^{i+1}{\cal J}_{s+2}\sim-&\Box (\partial_u\cdot\partial_u)^i{\cal J}_s+
(s+1)(s+d)(\partial_u\cdot\partial_u)^i{\cal J}_s\\
\notag
+\; &\sum_{k=0}^i(\partial_u\cdot\partial_u)^{i-k}(\Delta_+-\Delta_-)^2{\cal J}_{s-2k}
\frac{i!}{(i-k)!}4^k \left((d+1)/2+s-i-k\right)_k\\
\notag
-\;&\sum_{k=0}^{i-1}\left(\Box-(s-2i)(s-2i+d-1)\right)(\partial_u\cdot\partial_u)^{i-k-1}{\cal J}_{s-2k-2}\\
\label{appb2}
&\qquad\qquad\times\;
\frac{i!}{(i-k-1)!}4^{k+1} \left((d+1)/2+s-i-k\right)_{k+1}.
\end{align}
One notices that applying \eqref{appb2} to the right combination of 
$(\partial_u\cdot\partial_u)^{i+1}{\cal J}_{s+2}$ and $(\partial_u\cdot\partial_u)^{i}{\cal J}_{s}$,
the tails of terms on the right hand side cancel. To wit,
\begin{align}
\notag
(\partial_u\cdot\partial_u)^{i+1}&{\cal J}_{s+2}-2i(d+2s-2i-1)(\partial_u\cdot\partial_u)^{i}{\cal J}_{s}\\
\notag
&\sim\left((\Delta_+-\Delta_-)^2-\Box+(s+1)(s+d)\right)(\partial_u\cdot\partial_u)^{i}{\cal J}_{s} \\
\notag
&-2i\left(\Box-(s-2i)(s-2i+d-1)\right)(\partial_u\cdot\partial_u)^{i-1}{\cal J}_{s-2}\\
\label{appb3}
&-2i(d+2s-2i-1)\left(\Box-(s-1)(s+d-2)\right)(\partial_u\cdot\partial_u)^{i-1}{\cal J}_{s-2}.
\end{align}
From now we assume that both sides of \eqref{appb3} are integrated against the traceless and divergence's harmonic
function $\Omega_{s-2k}$. Then one can integrate by parts all $\Box$'s that appear
 on the right hand side of \eqref{appb3}, thus making them act on $\Omega_{s-2k}$.
Using \eqref{harmoniceq} and \eqref{massdimension} we obtain the following iterative 
equation
\begin{equation}
\label{appb4}
(\partial_u\cdot\partial_u)^{i+1}{\cal J}_{s+2}\sim f(i,s)(\partial_u\cdot\partial_u)^{i}{\cal J}_{s}+
g(i,s)(\partial_u\cdot\partial_u)^{i-1}{\cal J}_{s-2},
\end{equation}
where 
\begin{align*}
&f(i,s)=(2\Delta-d-1)(2\Delta-d+1)\\
&\qquad\qquad\qquad +\;\left(\nu^2+(\tfrac{d}{2}+(s-2i)+1)^2\right)+8i(\tfrac{d}{2}+(s-2i)+i),\\
&g(i,s)=4i(\tfrac{d}{2}+(s-2i)+i-1)\\
&\qquad\qquad\qquad\times\;\left(-\left(\nu^2+(\tfrac{d}{2}+(s-2i)+1)^2\right)-4(i-1)(\tfrac{d}{2}+(s-2i)+i)\right).
\end{align*}
As ``boundary conditions'' we take $(\partial_u\cdot\partial_u)^{-1}{\cal J}_{s-2k-2}\equiv 0$
and assume that ${\cal J}_{s-2k}$ is given. Then the iterative equation \eqref{appb4}
allows to express $(\partial_u\cdot\partial_u)^{k}{\cal J}_{s}$ in terms of  ${\cal J}_{s-2k}$.
The result is
\begin{eqnarray*}
(\partial_u\cdot\partial_u)^{k}{\cal J}_{s}\sim \tau_{s,k}\;{\cal J}_{s-2k}, 
\end{eqnarray*}
where
\begin{eqnarray}
\notag
\tau_{s,k}\left(\nu\right)=\sum_{m=0}^k 
\frac{2^{2k}\cdot k!}{m!(k-m)!}
(\Delta-\tfrac{d}{2}-k+m+1/2)_{k-m}\\
\notag
\times\left(\frac{\tfrac{d}{2}+s-2m+1+i\nu}{2}\right)_{m}\left(\frac{\tfrac{d}{2}+s-2m+1-i\nu}{2}\right)_{m}.
\end{eqnarray}



\providecommand{\href}[2]{#2}\begingroup\raggedright\endgroup

\end{document}